\documentclass[12pt]{iopart}

\expandafter\let\csname equation*\endcsname\relax 
\expandafter\let\csname endequation*\endcsname\relax 
\usepackage{mathrsfs}
\usepackage{amsmath,amsfonts,amssymb}

\usepackage{iopams}
\usepackage{graphicx,psfrag}
\usepackage{multirow}
\usepackage{comment}
\usepackage{hyperref}
\usepackage{color}
\usepackage{longtable}

\def\GMc2{{\rm G M_{\odot} c^{-2}}}

\def\sf{{\rm SF}}

\usepackage{color}
\definecolor{cyan}{rgb}{0,0.9,0.9}
\definecolor{orange}{rgb}{0.9,0.5,0}
\definecolor{magenta}{rgb}{1,0,1}
\definecolor{purple}{rgb}{0.8,0.4,0.8}
\definecolor{gray}{rgb}{0.8242,0.8242,0.8242}

\begin{document}


\title[Dynamical Ejecta and Electromagnetic Counterparts of BNS mergers]
       {Modeling dynamical ejecta from binary neutron star mergers
       and implications for electromagnetic counterparts}       
       

\author{Tim Dietrich$^1$, Maximiliano Ujevic$^2$}

\address{$^1$ Max Planck Institute for Gravitational Physics, Albert Einstein Institute, D-14476 Golm, Germany}
\address{$^2$Centro de Ci\^encias Naturais e Humanas, Universidade Federal do ABC, 09210-580, Santo Andr\'e, S\~ao Paulo, Brazil}
 
\date{\today}

\begin{abstract}
In addition to the emission of gravitational waves (GWs) 
the coalescence and merger of two neutron stars 
will produce a variety of electromagnetic (EM) signals. 
In this work we combine a large set of numerical 
relativity simulations performed by different groups 
and we present fits for the mass, 
kinetic energy, and the velocities of the dynamical 
ejected material. 
Additionally, we comment on the geometry and  
composition of the ejecta and 
discuss the influence of the stars' individual 
rotation. 

The derived fits can be used to 
approximate the luminosity and lightcurve of the kilonovae (macronovae)
and to estimate the main properties of the radio flares. 
This correlation between the binary parameters and the EM signals 
allows in case of a GW detection 
to approximate possible EM counterparts when 
first estimates of the masses are available.  
After a possible kilonovae observation our 
results could also be used to restrict the 
region of the parameter space which has to be covered 
by numerical relativity simulations.
\end{abstract}

\pacs{
  04.25.D-,     
  04.30.Db,   
  95.30.Sf,     
  95.30.Lz,   
  97.60.Jd      
}

\maketitle

\section{Introduction}
\label{sec:intro}

The first detections of coalescing binary black hole (BBH) 
systems~\cite{Abbott:2016blz,Abbott:2016nmj} inaugurated the field of 
gravitational wave (GW) astronomy. 
Beside BBHs, binary neutron stars (BNS) are one of the expected 
sources for future GW detections~\cite{Aasi:2013wya,Abbott:2016ymx}. 
In contrast to BBH mergers, it is expected that BNS mergers 
produce electromagnetic (EM) signals, 
as kilonovae (also called macronovae), 
radio flares or short gamma-ray bursts (SGRBs). 
While SGRBs are powered by collimated highly relativistic 
outflows, e.g.,~\cite{Paczynski:1986px,Eichler:1989ve,Soderberg:2006bn}, 
kilonovae are transient emissions in the optical or
near-infrared band, e.g.,~\cite{Tanvir:2013pia,Yang:2015pha,Jin:2016pnm}, 
produced by the radioactive decay of r-process nuclei 
in the neutron-rich material ejected during the merger. 
Additionally, mildly and sub- relativistic outflows can generate 
synchrotron radiation (radio flares) even years after the merger 
of the two neutron stars, see e.g.,~\cite{Nakar:2011cw}.   

One possibility to study BNS mergers are 
numerical relativity (NR) simulations. Those simulations 
allow to describe the system even beyond the merger of the two stars solving 
Einsteins field equations. Over the last years more microphysical descriptions 
have been included, e.g., realistic equation of states (EOSs), 
neutrino transport, magnetic fields. 
It also has become a common approach to extract information from
NR simulations about the unbound material ejected from the system 
and use these information to estimate possible EM counterparts.
However, the computation of ejecta and lightcurves is still challenging. 
While current state-of-the art numerical simulations cover 
the last $10-20$ orbits before and up to $\sim 50$ms after the merger, 
it is computationally too expensive to study the dynamical ejected material 
longer than a fraction of a second. 
But, it is possible to use relativistic simulations 
as initial conditions and either assume free expansion of the 
ejecta material, e.g.,~\cite{Goriely:2011vg}, 
evolution on a fixed spacetime background, e.g.,~\cite{Rosswog:2013kqa,Grossman:2013lqa},
or use radiative transfer Monte-Carlo simulations, e.g.,~\cite{Tanaka:2013ana,Hotokezaka:2013kza}. 
Our work is complementary to most previous studies, we will use a 
large set of numerical relativity data obtained from 
different groups to derive phenomenological fits 
relating the binary parameters to the ejecta properties. 
Knowing the basic properties of the 
ejecta allows to give estimates 
on the expected kilonovae and radio flares. 

In general, the time between a GW detection and the observation 
of the corresponding kilonovae (about a few days) is not long enough 
to perform full NR simulations which have typical run times of weeks to months. 
Therefore, NR simulations can only be used for comparison once GW and EM observations 
finish. The advantage of the phenomenological model proposed in this article 
is that even before the EM follow up observations start first estimates of the kilonovae
properties can be given.
Furthermore, after the kilonovae has been detected, the model can be used 
to reduce the part of the BNS parameter space which 
has to be covered by full NR simulations.

\section{Employed Dataset}
\label{sec:data}

\begin{figure}[t]
\begin{center}
  \includegraphics[width=0.8\textwidth]{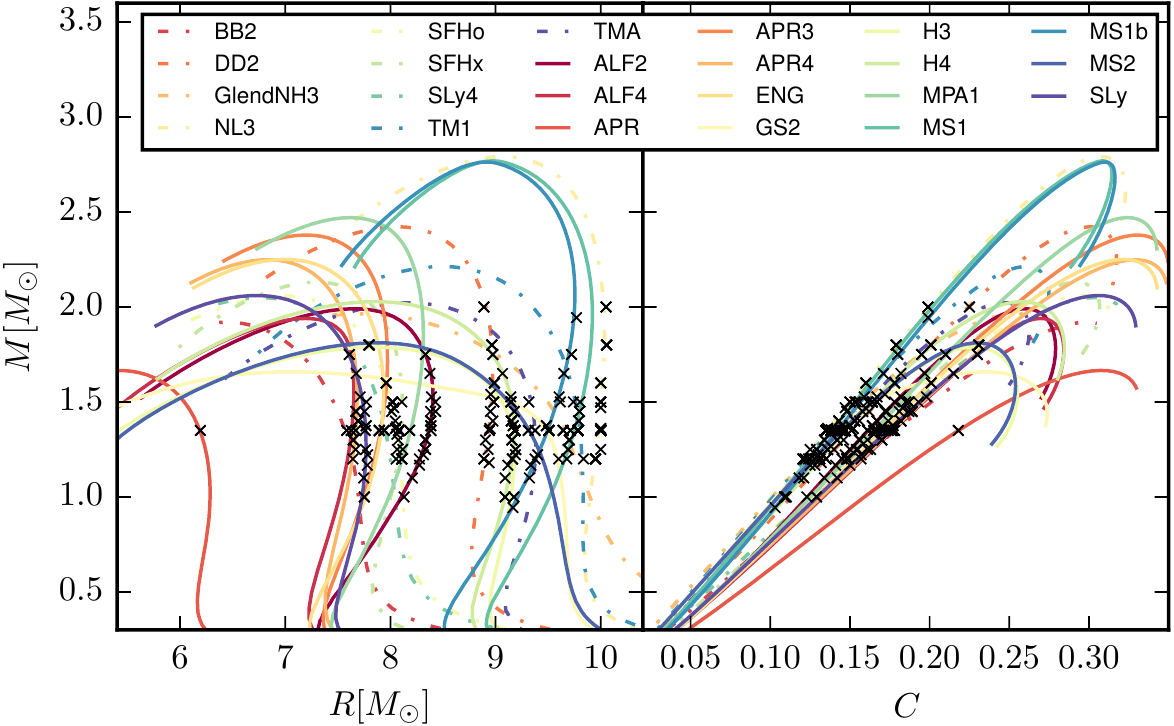}
  \caption{Mass vs.~radius relations (left) and mass vs.~compactness relations (right) 
           for all EOSs used in this work. 
           Tabulated EOSs are marked with dashed lines, piecewise polytropes with solid lines. 
           The markers refer to configurations employed in this work.}
  \label{fig:EOS}
  \end{center}
\end{figure}

Over the last years numerical relativity (NR) has made a tremendous progress and a large number of 
groups have studied the merger process of BNSs, see e.g.,~\cite{Faber:2012rw,Baiotti:2016qnr} 
and references therein. 
Despite the computation of the emitted GW signal, the investigation of 
ejected material and EM counterparts went into the focus of research.

Combining published work from different groups enables us to obtain 
an NR catalog to derive fitting formulas 
for important ejecta quantities. In this article we use results 
from~\cite{Hotokezaka:2012ze,Bauswein:2013yna,Dietrich:2015iva,Lehner:2016lxy,Sekiguchi:2016bjd,Dietrich:2016hky}, 
where the mass, kinetic energy, and velocity of the ejecta are reported.
The data set combines results based on grid structured 
codes~\cite{Hotokezaka:2012ze,Dietrich:2015iva,Lehner:2016lxy,Sekiguchi:2016bjd,Dietrich:2016hky}
with results employing a SPH code~\cite{Bauswein:2013yna} 
under conformal flatness approximation 
and it includes simplifies EOSs, 
tabulated EOS as well as simulations with and without neutrino treatment. 
In total $172$ simulations have been considered. 

Although simulation techniques are continuously improved and 
higher accuracy is achieved, 
the characterization of ejecta is still challenging and results have to be assigned with 
large uncertainties.
Considering the accuracy of the NR data points, 
quantities as the mass and kinetic energy have uncertainties which range between $\sim 10\%$
up to even $\sim 100\%$, see e.g.,~appendix~A of~\cite{Hotokezaka:2012ze} 
and table~III of~\cite{Dietrich:2016hky}, 
where multiple resolutions have been employed. 
In general one finds that the fractional 
uncertainty is larger for lower massive ejecta. 

In addition to the uncertainty of the results employing the same numerical code also 
differences between different implementations/codes exist. 
For some cases those discrepancies are quite large (up to a factor of $\sim 5$ in extreme cases) and 
they also depend on the implementation of thermal effects and if 
neutrino cooling or transport is included in the simulations. 
Those differences can produce systematic uncertainties. We try to minimize selection 
effects by including a large number of simulations produced by 
a variety of numerical codes. 
In the future crosschecks among different codes employing the same physical systems will be needed
for a better estimate of systematic errors. 

In our work, we restrict our analysis to dynamical ejecta. Ejecta produced 
after BH formation are not included, but will contribute to the 
total amount of ejecta and to the corresponding EM signals, 
see e.g.,~\cite{Metzger:2016pju}.
Thus, our results can be seen as lower bounds for the luminosity 
of EM observables.
Furthermore, while some of our data points were computed by NR 
simulations including neutrinos and tabulated EOSs, 
the effect of magnetic fields is not studied, 
although magnetic fields will influence the binary 
dynamics shortly around and after merger 
and lead to mass ejection by 
magnetic winds. 

The complete dataset is reported in table~\ref{tab:overview}, where a simulation number 
is assigned to every data point (first column).
In total we consider 23 different EOSs (shown in figure~\ref{fig:EOS}). 
Most EOSs are represented by a piecewise polytrope fitted to a zero-temperature EOS (straight lines), 
see e.g.,~\cite{Read:2009yp}. An additional thermal contribution 
to the pressure according to $p_{\rm th} = \rho \epsilon (\Gamma_{\rm th} - 1)$ 
is added for the evolution, where $\rho$ is the rest-mass density and $\epsilon$
the internal energy. The parameter
$\Gamma_{\rm th}$ is also reported in table~\ref{tab:overview}. 
Some simulations use full tabulated EOSs (dashed lines), 
which we denote as full in table~\ref{tab:overview}. 
Simulations with tabulated EOSs and neutrino treatment are denoted with fullN.
In addition to the parameters describing the binary, we report the mass of the 
ejected material $M_{\rm ej}$, the kinetic energy $T_{\rm ej}$, the average velocity
inside the orbital plane $v_\rho$, the average velocity perpendicular to the 
orbital plane $v_z$, and the total velocity $v_{\rm ej}$.   

\begin{small}
  \renewcommand{\arraystretch}{0.9}
  \begin{longtable}{ll|cccc|ccccc} 
  \caption{ \label{tab:overview} NR data used in this work. Columns refer to: 
    The data ID, cf.~e.g.,~figure~\ref{fig:Mej}, 
    mass of the first star $M_1$, 
    mass of the second star $M_2$, 
    $\Gamma_{\rm th}$ modeling thermal effects for piecewise polytropic EOS,
    ejecta mass $M_{\rm ej}$,
    kinetic energy of the ejecta $T_{\rm ej}$, 
    average velocity inside the orbital plane $v_{\rho}$, 
    average velocity perpendicular to the orbital plane $v_z$, 
    total average ejecta velocity $v_{\rm ej}$.
    In cases where $v_\rho$ and $v_z$ are given, we 
    estimate the total ejecta velocity as 
    $v_{\rm ej} = \sqrt{v_\rho^2+v_z^2}$. 
    Note that in~\cite{Sekiguchi:2016bjd}
    the ejecta velocity was estimated based on 
    $T_{\rm ej} = M_{\rm ej} v^2_{\rm ej}/2$, 
    consequently we use this relation to compute the kinetic energy
    not stated in~\cite{Sekiguchi:2016bjd}.}\\

    $\#$ & Ref & EOS & $M_1$     & $M_2$     & $\Gamma_{\rm th}$ & $M_{\rm ej}$       & $T_{\rm ej}$  & $v_{\rho}$ & $v_{z}$ & $v_{\rm ej}$\\ 
         &     &     &$[M_\odot]$&$[M_\odot$]&                   &$[10^{-3}M_\odot]$  & $[10^{50}$erg]& $[c]$    & $[c]$ & $[c]$ \\  
    \hline
1 & ALF2 & \cite{Dietrich:2016hky} & 1 & 1.75 & 1.75 & 36 & 12.69 & 0.18 & 0.03 & 0.18 \\
2 & ALF2 & \cite{Dietrich:2016hky} & 1.167 & 1.75 & 1.75 & 25 & 10.73 & 0.19 & 0.06 & 0.2 \\
3 & ALF2 & \cite{Dietrich:2016hky} & 1.1 & 1.65 & 1.75 & 24 & 7.5 & 0.17 & 0.07 & 0.18 \\
4 & ALF2 & \cite{Dietrich:2016hky} & 1 & 1.5 & 1.75 & 21 & 4.8 & 0.15 & 0.07 & 0.17 \\
5 & ALF2 & \cite{Dietrich:2016hky} & 1.222 & 1.527 & 1.75 & 7.5 & 3.93 & 0.17 & 0.12 & 0.21 \\
6 & ALF2 & \cite{Hotokezaka:2012ze} & 1.2 & 1.5 & 1.8 & 5.5 & 3 & 0.21 & 0.1 & 0.23 \\
7 & ALF2 & \cite{Hotokezaka:2012ze} & 1.25 & 1.45 & 1.8 & 3 & 1.5 & 0.2 & 0.1 & 0.22 \\
8 & ALF2 & \cite{Hotokezaka:2012ze} & 1.3 & 1.4 & 1.8 & 1.5 & 0.8 & 0.16 & 0.11 & 0.19 \\
9 & ALF2 & \cite{Hotokezaka:2012ze} & 1.4 & 1.4 & 1.8 & 2.5 & 1.5 & 0.21 & 0.13 & 0.25 \\
10 & ALF2 & \cite{Dietrich:2016hky} & 1.375 & 1.375 & 1.75 & 3.4 & 1.36 & 0.17 & 0.1 & 0.2 \\
11 & ALF2 & \cite{Hotokezaka:2012ze} & 1.35 & 1.35 & 1.8 & 2.5 & 1.5 & 0.22 & 0.12 & 0.25 \\
12 & ALF2 & \cite{Hotokezaka:2012ze} & 1.3 & 1.3 & 1.8 & 2 & 1 & 0.19 & 0.1 & 0.21 \\
13 & APR4 & \cite{Hotokezaka:2012ze} & 1.2 & 1.5 & 2 & 7.5 & 5.5 & 0.24 & 0.12 & 0.27 \\
14 & APR4 & \cite{Hotokezaka:2012ze} & 1.2 & 1.5 & 1.8 & 8 & 5.5 & 0.23 & 0.11 & 0.25 \\
15 & APR4 & \cite{Hotokezaka:2012ze} & 1.2 & 1.5 & 1.6 & 9 & 5 & 0.2 & 0.1 & 0.22 \\
16 & APR4 & \cite{Hotokezaka:2012ze} & 1.3 & 1.6 & 1.8 & 2 & 1.5 & 0.24 & 0.08 & 0.25 \\
17 & APR4 & \cite{Hotokezaka:2012ze} & 1.2 & 1.4 & 1.8 & 3 & 2 & 0.21 & 0.12 & 0.24 \\
18 & APR4 & \cite{Hotokezaka:2012ze} & 1.25 & 1.45 & 1.8 & 7 & 4.5 & 0.22 & 0.11 & 0.25 \\
19 & APR4 & \cite{Hotokezaka:2012ze} & 1.3 & 1.5 & 1.8 & 12 & 8.5 & 0.23 & 0.12 & 0.26 \\
20 & APR4 & \cite{Hotokezaka:2012ze} & 1.3 & 1.4 & 1.8 & 8 & 5 & 0.19 & 0.12 & 0.22 \\
21 & APR4 & \cite{Hotokezaka:2012ze} & 1.25 & 1.35 & 1.8 & 5 & 3 & 0.18 & 0.1 & 0.21 \\
22 & APR4 & \cite{Hotokezaka:2012ze} & 1.4 & 1.5 & 1.8 & 0.6 & 0.9 & 0.35 & 0.12 & 0.37 \\
23 & APR4 & \cite{Hotokezaka:2012ze} & 1.45 & 1.45 & 1.8 & 0.1 & 0.1 & 0.29 & 0.13 & 0.32 \\
24 & APR4 & \cite{Hotokezaka:2012ze} & 1.4 & 1.4 & 1.8 & 14 & 10 & 0.22 & 0.15 & 0.27 \\
25 & APR4 & \cite{Hotokezaka:2012ze} & 1.35 & 1.35 & 2 & 5 & 3 & 0.19 & 0.13 & 0.23 \\
26 & APR4 & \cite{Hotokezaka:2012ze} & 1.35 & 1.35 & 1.8 & 7 & 4 & 0.19 & 0.12 & 0.22 \\
27 & APR4 & \cite{Hotokezaka:2012ze} & 1.35 & 1.35 & 1.6 & 11 & 6 & 0.19 & 0.13 & 0.23 \\
28 & APR4 & \cite{Hotokezaka:2012ze} & 1.3 & 1.3 & 1.8 & 2 & 1 & 0.19 & 0.1 & 0.21 \\
29 & H4 & \cite{Dietrich:2016hky} & 1 & 1.75 & 1.75 & 40 & 12.51 & 0.17 & 0.02 & 0.17 \\
30 & H4 & \cite{Dietrich:2016hky} & 1.167 & 1.75 & 1.75 & 14 & 4.65 & 0.18 & 0.05 & 0.19 \\
31 & H4 & \cite{Dietrich:2016hky} & 1.1 & 1.65 & 1.75 & 17 & 4.83 & 0.17 & 0.04 & 0.17 \\
32 & H4 & \cite{Dietrich:2016hky} & 1 & 1.5 & 1.75 & 27 & 8.04 & 0.17 & 0.03 & 0.17 \\
33 & H4 & \cite{Dietrich:2016hky} & 1.222 & 1.527 & 1.75 & 6.6 & 3.04 & 0.18 & 0.11 & 0.21 \\
34 & H4 & \cite{Hotokezaka:2012ze} & 1.2 & 1.5 & 2 & 4 & 2 & 0.21 & 0.09 & 0.23 \\
35 & H4 & \cite{Hotokezaka:2012ze} & 1.2 & 1.5 & 1.8 & 3.5 & 2 & 0.21 & 0.09 & 0.23 \\
36 & H4 & \cite{Hotokezaka:2012ze} & 1.2 & 1.5 & 1.6 & 4.5 & 2 & 0.19 & 0.1 & 0.21 \\
37 & H4 & \cite{Hotokezaka:2012ze} & 1.2 & 1.4 & 1.8 & 2.5 & 1 & 0.19 & 0.1 & 0.21 \\
38 & H4 & \cite{Hotokezaka:2012ze} & 1.25 & 1.45 & 1.8 & 2 & 1.5 & 0.19 & 0.1 & 0.21 \\
39 & H4 & \cite{Hotokezaka:2012ze} & 1.3 & 1.5 & 1.8 & 3 & 2 & 0.19 & 0.1 & 0.21 \\
40 & H4 & \cite{Hotokezaka:2012ze} & 1.3 & 1.4 & 1.8 & 0.7 & 0.4 & 0.18 & 0.1 & 0.21 \\
41 & H4 & \cite{Hotokezaka:2012ze} & 1.25 & 1.35 & 1.8 & 0.6 & 0.3 & 0.18 & 0.1 & 0.21 \\
42 & H4 & \cite{Hotokezaka:2012ze} & 1.4 & 1.4 & 1.8 & 0.3 & 0.2 & 0.17 & 0.13 & 0.21 \\
43 & H4 & \cite{Dietrich:2016hky} & 1.375 & 1.375 & 1.75 & 3.4 & 1.59 & 0.19 & 0.1 & 0.21 \\
44 & H4 & \cite{Hotokezaka:2012ze} & 1.35 & 1.35 & 2 & 0.4 & 0.2 & 0.2 & 0.1 & 0.22 \\
45 & H4 & \cite{Hotokezaka:2012ze} & 1.35 & 1.35 & 1.8 & 0.5 & 0.2 & 0.19 & 0.11 & 0.22 \\
46 & H4 & \cite{Hotokezaka:2012ze} & 1.35 & 1.35 & 1.6 & 0.7 & 0.4 & 0.21 & 0.11 & 0.24 \\
47 & H4 & \cite{Hotokezaka:2012ze} & 1.3 & 1.3 & 1.8 & 0.3 & 0.1 & 0.16 & 0.1 & 0.19 \\
48 & MS1 & \cite{Hotokezaka:2012ze} & 1.2 & 1.5 & 1.8 & 3.5 & 1.5 & 0.19 & 0.1 & 0.21 \\
49 & MS1 & \cite{Hotokezaka:2012ze} & 1.25 & 1.45 & 1.8 & 1.5 & 0.8 & 0.19 & 0.11 & 0.22 \\
50 & MS1 & \cite{Hotokezaka:2012ze} & 1.3 & 1.4 & 1.8 & 0.6 & 0.2 & 0.17 & 0.09 & 0.19 \\
51 & MS1 & \cite{Hotokezaka:2012ze} & 1.4 & 1.4 & 1.8 & 0.6 & 0.2 & 0.13 & 0.09 & 0.16 \\
52 & MS1 & \cite{Hotokezaka:2012ze} & 1.35 & 1.35 & 1.8 & 1.5 & 0.6 & 0.14 & 0.08 & 0.16 \\
53 & MS1 & \cite{Hotokezaka:2012ze} & 1.3 & 1.3 & 1.8 & 1.5 & 0.5 & 0.15 & 0.08 & 0.17 \\
54 & MS1b & \cite{Dietrich:2016hky} & 0.944 & 1.944 & 1.75 & 65 & 21.45 & 0.18 & 0.02 & 0.18 \\
55 & MS1b & \cite{Dietrich:2016hky} & 1 & 1.75 & 1.75 & 49 & 15.19 & 0.17 & 0.03 & 0.17 \\
56 & MS1b & \cite{Dietrich:2016hky} & 1.167 & 1.75 & 1.75 & 24 & 7.69 & 0.18 & 0.05 & 0.19 \\
57 & MS1b & \cite{Dietrich:2016hky} & 1.1 & 1.65 & 1.75 & 26 & 7.33 & 0.17 & 0.04 & 0.17 \\
58 & MS1b & \cite{Dietrich:2016hky} & 1 & 1.5 & 1.75 & 32 & 7.87 & 0.16 & 0.03 & 0.16 \\
59 & MS1b & \cite{Dietrich:2016hky} & 1.222 & 1.527 & 1.75 & 4.8 & 1.64 & 0.15 & 0.11 & 0.19 \\ 
60 & MS1b & \cite{Dietrich:2016hky} & 1.375 & 1.375 & 1.75 & 2.3 & 0.39 & 0.13 & 0.06 & 0.14 \\
61 & SLy & \cite{Dietrich:2016hky} & 1 & 1.75 & 1.75 & 24 & 8.94 & 0.19 & 0.03 & 0.19 \\
62 & SLy & \cite{Dietrich:2016hky} & 1.167 & 1.75 & 1.75 & 6.5 & 5.54 & 0.25 & 0.11 & 0.27 \\
63 & SLy & \cite{Dietrich:2016hky} & 1.1 & 1.65 & 1.75 & 16 & 7.69 & 0.19 & 0.11 & 0.22 \\
64 & SLy & \cite{Dietrich:2016hky} & 1 & 1.5 & 1.75 & 18 & 9.12 & 0.19 & 0.12 & 0.22 \\
65 & SLy & \cite{Dietrich:2016hky} & 1.222 & 1.527 & 1.75 & 18 & 8.4 & 0.16 & 0.11 & 0.19 \\
66 & SLy & \cite{Dietrich:2016hky} & 1.375 & 1.375 & 1.75 & 16 & 4.83 & 0.17 & 0.1 & 0.2 \\
67 & ALF2 & \cite{Dietrich:2015iva} & 1.25 & 1.45 & 1.75 & 3.9 & 0.8 & - & - & 0.15 \\ 
68 & ALF2 & \cite{Bauswein:2013yna} & 1.35 & 1.35 & 2 & 3.8 & 3.36 & - & - & 0.28 \\
69 & ALF2 & \cite{Dietrich:2015iva} & 1.35 & 1.35 & 1.75 & 3.5 & 0.7 & - & - & 0.15 \\
70 & ALF2 & \cite{Bauswein:2013yna} & 1.35 & 1.35 & 1.5 & 4.49 & 3.8 & - & - & 0.27\\
71 & ALF4 & \cite{Bauswein:2013yna} & 1.35 & 1.35 & 2 & 5.7 & 6.07 & - & - & 0.3\\
72 & ALF4 & \cite{Bauswein:2013yna} & 1.35 & 1.35 & 1.5 & 7.4 & 7.65 & - & - & 0.29\\
73 & APR & \cite{Bauswein:2013yna} & 1.35 & 1.35 & 2 & 5.96 & 6.37 & - & - & 0.31\\
74 & APR & \cite{Bauswein:2013yna} & 1.35 & 1.35 & 1.5 & 7.38 & 7.9 & - & - & 0.3\\
75 & APR3 & \cite{Bauswein:2013yna} & 1.35 & 1.35 & 2 & 4.65 & 4.69 & - & - & 0.3\\
76 & APR3 & \cite{Bauswein:2013yna} & 1.35 & 1.35 & 1.5 & 6.15 & 5.5 & - & - & 0.27\\
77 & DD2 & \cite{Bauswein:2013yna} & 1.2 & 1.8 & full & 17.08 & 6.72 & - & - & 0.17\\
78 & DD2 & \cite{Bauswein:2013yna} & 1.35 & 2 & full & 6.41 & 9.64 & - & - & 0.31\\
79 & DD2 & \cite{Bauswein:2013yna} & 1.35 & 1.8 & full & 14.85 & 9.48 & - & - & 0.21\\
80 & DD2 & \cite{Bauswein:2013yna} & 1.2 & 1.6 & full & 10.9 & 6.39 & - & - & 0.2\\
81 & DD2 & \cite{Lehner:2016lxy} & 1.18 & 1.54 & fullN & 1.3 & 0.76 & - & - & 0.3\\
82 & DD2 & \cite{Bauswein:2013yna} & 1.2 & 1.5 & full & 8.79 & 4.97 & - & - & 0.2\\
83 & DD2 & \cite{Bauswein:2013yna} & 1.5 & 1.8 & full & 18.84 & 15.52 & - & - & 0.25\\
84 & DD2 & \cite{Lehner:2016lxy} & 1.25 & 1.47 & fullN & 0.42 & 0.29 & - & - & 0.3\\
85 & DD2 & \cite{Sekiguchi:2016bjd} & 1.25 & 1.45 & fullN & 5 & 1.61 & - & - & 0.19\\
86 & DD2 & \cite{Bauswein:2013yna} & 1.2 & 1.35 & full & 3.17 & 2.06 & - & - & 0.2\\
87 & DD2 & \cite{Bauswein:2013yna} & 1.35 & 1.5 & full & 3.57 & 3.13 & - & - & 0.25\\
88 & DD2 & \cite{Sekiguchi:2016bjd} & 1.3 & 1.4 & fullN & 3 & 0.87 & - & - & 0.18\\
89 & DD2 & \cite{Bauswein:2013yna} & 2 & 2 & full & 0.25 & 0.25 & - & - & 0.25\\
90 & DD2 & \cite{Bauswein:2013yna} & 1.8 & 1.8 & full & 1.37 & 1.63 & - & - & 0.26\\
91 & DD2 & \cite{Bauswein:2013yna} & 1.6 & 1.6 & full & 7.8 & 7.4 & - & - & 0.27\\
92 & DD2 & \cite{Bauswein:2013yna} & 1.5 & 1.5 & full & 5.38 & 4.66 & - & - & 0.26\\
93 & DD2 & \cite{Lehner:2016lxy} & 1.36 & 1.36 & fullN & 0.43 & 0.31 & - & - & 0.3\\
94 & DD2 & \cite{Bauswein:2013yna} & 1.35 & 1.35 & 2 & 2.57 & 3.31 & - & - & 0.34\\
95 & DD2 & \cite{Bauswein:2013yna} & 1.35 & 1.35 & 1.8 & 2.26 & 2.61 & - & - & 0.32\\
96 & DD2 & \cite{Bauswein:2013yna} & 1.35 & 1.35 & 1.5 & 2.72 & 2.9 & - & - & 0.3\\
97 & DD2 & \cite{Bauswein:2013yna} & 1.35 & 1.35 & full & 3.07 & 2.18 & - & - & 0.22\\
98 & DD2 & \cite{Sekiguchi:2016bjd} & 1.35 & 1.35 & fullN & 2 & 0.46 & - & - & 0.16\\
99 & DD2 & \cite{Bauswein:2013yna} & 1.2 & 1.2 & full & 3.09 & 1.37 & - & - & 0.17\\
100 & ENG & \cite{Bauswein:2013yna} & 1.35 & 1.35 & 2 & 5.29 & 5.01 & - & - & 0.29\\
101 & ENG & \cite{Bauswein:2013yna} & 1.35 & 1.35 & 1.5 & 6.32 & 5.3 & - & - & 0.26\\
102 & Glenh3 & \cite{Bauswein:2013yna} & 1.35 & 1.35 & 2 & 1.08 & 0.62 & - & - & 0.23\\
103 & Glenh3 & \cite{Bauswein:2013yna} & 1.35 & 1.35 & 1.5 & 1.69 & 0.9 & - & - & 0.22\\
104 & GS2 & \cite{Bauswein:2013yna} & 1.2 & 1.5 & full & 10.69 & 6.14 & - & - & 0.18\\
105 & GS2 & \cite{Bauswein:2013yna} & 1.35 & 1.35 & full & 2.74 & 2.16 & - & - & 0.19\\
106 & H3 & \cite{Bauswein:2013yna} & 1.35 & 1.35 & 2 & 1.43 & 1.15 & - & - & 0.27\\
107 & H4 & \cite{Dietrich:2015iva} & 1.25 & 1.45 & 1.75 & 6 & 2.8 & - & - & 0.23\\
108 & H4 & \cite{Bauswein:2013yna} & 1.35 & 1.35 & 2 & 1.28 & 1.09 & - & - & 0.27\\
109 & H4 & \cite{Dietrich:2015iva} & 1.35 & 1.35 & 1.75 & 0.6 & 0.5 & - & - & 0.3\\
110 & H4 & \cite{Bauswein:2013yna} & 1.35 & 1.35 & 1.5 & 1.93 & 1.64 & - & - & 0.27\\
111 & MPA1 & \cite{Bauswein:2013yna} & 1.35 & 1.35 & 2 & 3.64 & 3.6 & - & - & 0.3\\
112 & MPA1 & \cite{Bauswein:2013yna} & 1.35 & 1.35 & 1.5 & 4.48 & 4.35 & - & - & 0.29\\
113 & MS1 & \cite{Dietrich:2015iva} & 1.25 & 1.45 & 1.75 & 5.8 & 1.2 & - & - & 0.15\\
114 & MS1 & \cite{Bauswein:2013yna} & 1.35 & 1.35 & 2 & 1.17 & 0.98 & - & - & 0.27\\
115 & MS1 & \cite{Dietrich:2015iva} & 1.35 & 1.35 & 1.75 & 0.7 & 0.2 & - & - & 0.18\\
116 & MS1 & \cite{Bauswein:2013yna} & 1.35 & 1.35 & 1.5 & 2.38 & 1.19 & - & - & 0.21\\
117 & MS1b & \cite{Bauswein:2013yna} & 1.35 & 1.35 & 2 & 1.67 & 1.26 & - & - & 0.25\\
118 & MS1b & \cite{Bauswein:2013yna} & 1.35 & 1.35 & 1.5 & 3.64 & 1.85 & - & - & 0.21\\
119 & MS2 & \cite{Bauswein:2013yna} & 1.35 & 1.35 & 2 & 0.81 & 0.65 & - & - & 0.26\\
120 & NL3 & \cite{Bauswein:2013yna} & 1.2 & 1.8 & full & 15.68 & 5.75 & - & - & 0.15\\
121 & NL3 & \cite{Bauswein:2013yna} & 1.35 & 2 & full & 12.85 & 7.62 & - & - & 0.2\\
122 & NL3 & \cite{Bauswein:2013yna} & 1.35 & 1.8 & full & 18.81 & 11.31 & - & - & 0.21\\
123 & NL3 & \cite{Bauswein:2013yna} & 1.2 & 1.6 & full & 9.96 & 5.57 & - & - & 0.19\\
124 & NL3 & \cite{Bauswein:2013yna} & 1.2 & 1.5 & full & 7.95 & 4.5 & - & - & 0.19\\
125 & NL3 & \cite{Bauswein:2013yna} & 1.5 & 1.8 & full & 8.1 & 4.94 & - & - & 0.21\\
126 & NL3 & \cite{Lehner:2016lxy} & 1.25 & 1.47 & fullN & 2.3 & 1.22 & - & - & 0.25\\
127 & NL3 & \cite{Bauswein:2013yna} & 1.35 & 1.5 & full & 2.72 & 2.25 & - & - & 0.24\\
128 & NL3 & \cite{Bauswein:2013yna} & 1.2 & 1.35 & full & 4.25 & 2.74 & - & - & 0.21\\
129 & NL3 & \cite{Bauswein:2013yna} & 2 & 2 & full & 1.91 & 2.18 & - & - & 0.29\\
130 & NL3 & \cite{Bauswein:2013yna} & 1.8 & 1.8 & full & 9.08 & 7.25 & - & - & 0.24\\
131 & NL3 & \cite{Bauswein:2013yna} & 1.6 & 1.6 & full & 3.74 & 2.59 & - & - & 0.22\\
132 & NL3 & \cite{Bauswein:2013yna} & 1.5 & 1.5 & full & 1.7 & 1.04 & - & - & 0.2\\
133 & NL3 & \cite{Lehner:2016lxy} & 1.36 & 1.36 & fullN & 0.015 & 0.01 & - & - & 0.45\\
134 & NL3 & \cite{Bauswein:2013yna} & 1.35 & 1.35 & 2 & 1.57 & 2.03 & - & - & 0.34\\
135 & NL3 & \cite{Bauswein:2013yna} & 1.35 & 1.35 & 1.8 & 1.6 & 2.99 & - & - & 0.32\\
136 & NL3 & \cite{Bauswein:2013yna} & 1.35 & 1.35 & 1.5 & 1.86 & 1.98 & - & - & 0.3\\
137 & NL3 & \cite{Bauswein:2013yna} & 1.35 & 1.35 & full & 2.09 & 0.98 & - & - & 0.18\\
138 & NL3 & \cite{Bauswein:2013yna} & 1.2 & 1.2 & full & 2.15 & 0.91 & - & - & 0.17\\
139 & SFHo & \cite{Bauswein:2013yna} & 1.2 & 1.8 & full & 5.78 & 10.08 & - & - & 0.34\\
140 & SFHo & \cite{Bauswein:2013yna} & 1.35 & 1.8 & full & 11.76 & 16.22 & - & - & 0.31\\
141 & SFHo & \cite{Bauswein:2013yna} & 1.2 & 1.6 & full & 16.91 & 11.1 & - & - & 0.21\\
142 & SFHo & \cite{Bauswein:2013yna} & 1.2 & 1.5 & full & 13.39 & 8.94 & - & - & 0.22\\
143 & SFHo & \cite{Bauswein:2013yna} & 1.5 & 1.8 & full & 6.34 & 14.4 & - & - & 0.42\\
144 & SFHo & \cite{Lehner:2016lxy} & 1.25 & 1.47 & fullN & 2.2 & 1.8 & - & - & 0.25\\
145 & SFHo & \cite{Sekiguchi:2016bjd} & 1.25 & 1.45 & fullN & 11 & 5.66 & - & - & 0.24\\
146 & SFHo & \cite{Bauswein:2013yna} & 1.2 & 1.35 & full & 5.44 & 3.86 & - & - & 0.22\\
147 & SFHo & \cite{Bauswein:2013yna} & 1.35 & 1.5 & full & 18.73 & 13.34 & - & - & 0.23\\
148 & SFHo & \cite{Sekiguchi:2016bjd} & 1.3 & 1.4 & fullN & 6 & 2.15 & - & - & 0.2\\
149 & SFHo & \cite{Sekiguchi:2016bjd} & 1.33 & 1.37 & fullN & 9 & 3.55 & - & - & 0.21\\
150 & SFHo & \cite{Bauswein:2013yna} & 1.8 & 1.8 & full & 0.17 & 0.24 & - & - & 0.29\\
151 & SFHo & \cite{Bauswein:2013yna} & 1.6 & 1.6 & full & 1.13 & 1 & - & - & 0.21\\
152 & SFHo & \cite{Bauswein:2013yna} & 1.5 & 1.5 & full & 4.1 & 4.13 & - & - & 0.27\\
153 & SFHo & \cite{Lehner:2016lxy} & 1.36 & 1.36 & fullN & 3.4 & 1.8 & - & - & 0.25\\
154 & SFHo & \cite{Bauswein:2013yna} & 1.35 & 1.35 & 2 & 2.96 & 3.37 & - & - & 0.32\\
155 & SFHo & \cite{Bauswein:2013yna} & 1.35 & 1.35 & 1.8 & 3.26 & 4.18 & - & - & 0.34\\
156 & SFHo & \cite{Bauswein:2013yna} & 1.35 & 1.35 & 1.5 & 3.82 & 4.14 & - & - & 0.3\\
157 & SFHo & \cite{Bauswein:2013yna} & 1.35 & 1.35 & full & 4.83 & 3.61 & - & - & 0.23\\
158 & SFHo & \cite{Sekiguchi:2016bjd} & 1.35 & 1.35 & fullN & 11 & 4.76 & - & - & 0.22\\
159 & SFHo & \cite{Bauswein:2013yna} & 1.2 & 1.2 & full & 1.88 & 1.26 & - & - & 0.21\\
160 & SFHx & \cite{Bauswein:2013yna} & 1.2 & 1.5 & full & 14.67 & 7.91 & - & - & 0.19\\
161 & SFHx & \cite{Bauswein:2013yna} & 1.35 & 1.35 & full & 6.16 & 4.36 & - & - & 0.22\\
162 & SLy & \cite{Dietrich:2015iva} & 1.25 & 1.45 & 1.75 & 6.5 & 5.1 & - & - & 0.3\\
163 & SLy & \cite{Dietrich:2015iva} & 1.35 & 1.35 & 1.75 & 12.2 & 7.1 & - & - & 0.26\\
164 & SLy4 & \cite{Bauswein:2013yna} & 1.35 & 1.35 & 2 & 3.99 & 3.75 & - & - & 0.29\\
165 & SLy4 & \cite{Bauswein:2013yna} & 1.35 & 1.35 & 1.5 & 6.4 & 5.53 & - & - & 0.27\\
166 & TM1 & \cite{Bauswein:2013yna} & 1.2 & 1.5 & full & 8.66 & 3.94 & - & - & 0.17\\
167 & TM1 & \cite{Bauswein:2013yna} & 1.35 & 1.35 & 2 & 1.37 & 2.02 & - & - & 0.36\\
168 & TM1 & \cite{Bauswein:2013yna} & 1.35 & 1.35 & 1.8 & 1.33 & 1.77 & - & - & 0.34\\
169 & TM1 & \cite{Bauswein:2013yna} & 1.35 & 1.35 & 1.5 & 1.53 & 1.86 & - & - & 0.32\\
170 & TM1 & \cite{Bauswein:2013yna} & 1.35 & 1.35 & full & 1.67 & 0.74 & - & - & 0.16\\
171 & TMA & \cite{Bauswein:2013yna} & 1.2 & 1.5 & full & 10.21 & 6.4 & - & - & 0.2\\
172 & TMA & \cite{Bauswein:2013yna} & 1.35 & 1.35 & full & 2.05 & 1.19 & - & - & 0.18\\
\end{longtable}
\end{small}
\section{Ejecta properties}
\label{sec:fits}

\subsection{Ejecta mass}

Considering EM signals from BNS mergers, one of the most important quantities
influencing the luminosity of kilonovae and radio flares is the mass of 
the material ejected from the system. 
The authors in \cite{Foucart:2012nc,Kawaguchi:2016ana} proposed
fitting formulas for the disk and ejecta mass for BHNS systems. 
To our knowledge no fit for the mass of the ejected material for BNS mergers exists to date.

Our fitting formula
\begin{equation}
 \frac{M_{\rm ej}^{\rm fit}}{10^{-3}M_\odot}  = \left[ a \left(\frac{M_2}{M_1}\right)^{1/3} \left(\frac{1-2 C_1}{C_1}\right)+ 
                          b \left(\frac{M_2}{M_1}\right)^{n} + c \left(1 - \frac{M_1}{M^*_{1}}\right) \right] M_1^* + (1\leftrightarrow 2) + d. \label{eq:Mej_fit}
\end{equation}
is an extension of the work done for BHNS systems to a system consisting of two neutron stars. 
We denote the mass in isolation of the i-th star as $M_i$,  
the baryonic mass as $M_i^*$, and the compactness as $C_i$.
Let us emphasize that although it has been shown that for BNS mergers 
a significant part of the ejecta is produced by shocks, 
e.g.,~\cite{Hotokezaka:2012ze}, \eqref{eq:Mej_fit} gives a robust estimate 
for the ejecta for almost all considered configurations. 
For our data we obtain the following fitting parameters:
\begin{equation}
 a = -1.35695 , \quad b = 6.11252  ,\quad  c =  -49.43355 ,\quad  d = 16.1144,\quad  n = -2.5484. \label{eq:par:Mej}
\end{equation}

\begin{figure}[t]
  \begin{center}
  \includegraphics[width=1\textwidth]{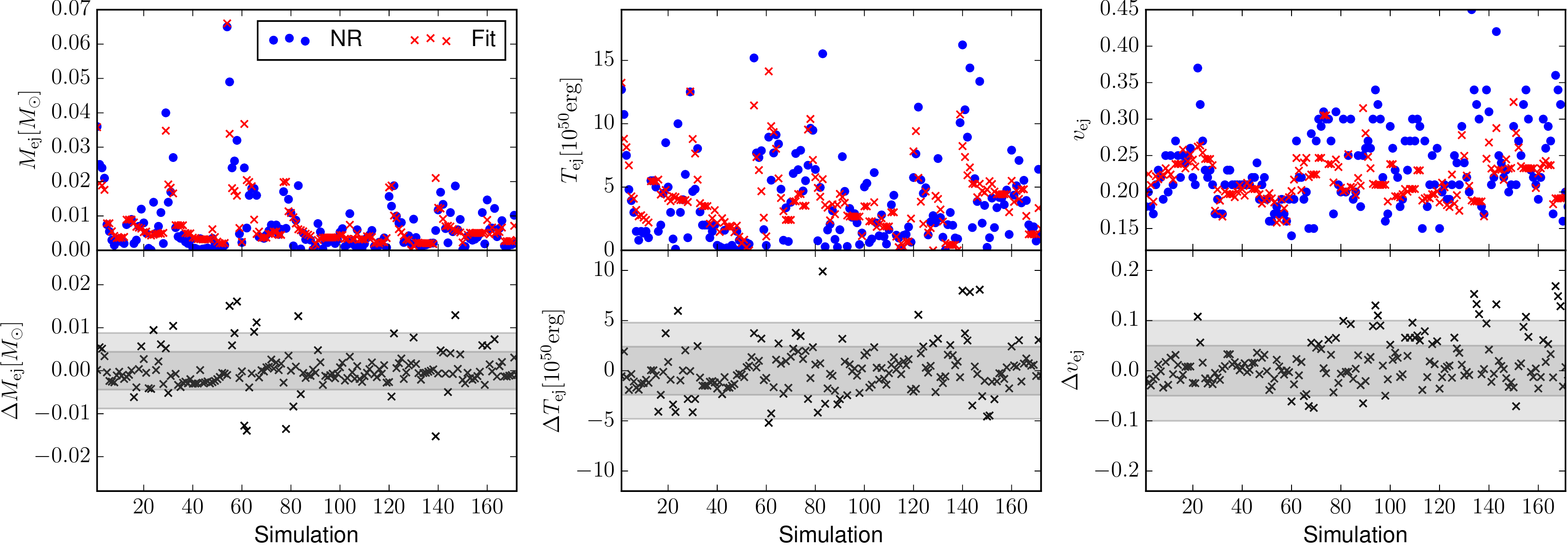}
  \caption{From left to right: ejecta mass $M_{\rm ej}$, kinetic energy of the ejecta $T_{\rm ej}$, and 
           velocity of the ejecta $v_{\rm ej}$. 
           The top panels show the NR data and the results obtained by our phenomenological fits. 
           The bottom panels show the absolute difference between the fit and the NR data, as shaded regions 
           we also include the 1-$\sigma$ and 2-$\sigma$ confidence interval.}
  \label{fig:Mej}
  \end{center}
\end{figure}

The left panels of figure~\ref{fig:Mej} show our results for the ejecta mass. 
In the upper panel we present $M_{\rm ej}$ 
for the numerical simulation (blue circles) and 
for our fitting formula $M_{\rm ej}^{\rm fit}$ (red crosses). 
Both quantities are plotted as a function of the simulation-ID
introduced in table~\ref{tab:overview}. 
The bottom panel shows the absolute residual $\Delta M_{\rm ej} = 
M_{\rm ej}^{\rm fit}-M_{\rm ej}$. We include as shaded regions the 
1$\sigma$ ($\Delta M_{\rm ej}^{1\sigma} = 4.4 \times 10^{-3}M_\odot$)
and $2\sigma$ confidence intervals. 
Our model function has an average residual of
$\Delta \bar{M}_{\rm ej} = 2.9\times10^{-3}M_\odot$, 
which corresponds to a fractional error of $\sim 72\%$.

Overall, because of the difficulties computing the 
ejecta properties, see section~\ref{sec:data}, 
$\Delta \bar{M}_{\rm ej}$ is of the same order as 
the numerical uncertainty of the NR data points and therefore can 
be considered as a possible estimate. 

\begin{figure}[t]
  \begin{center}
  \includegraphics[width=1\textwidth]{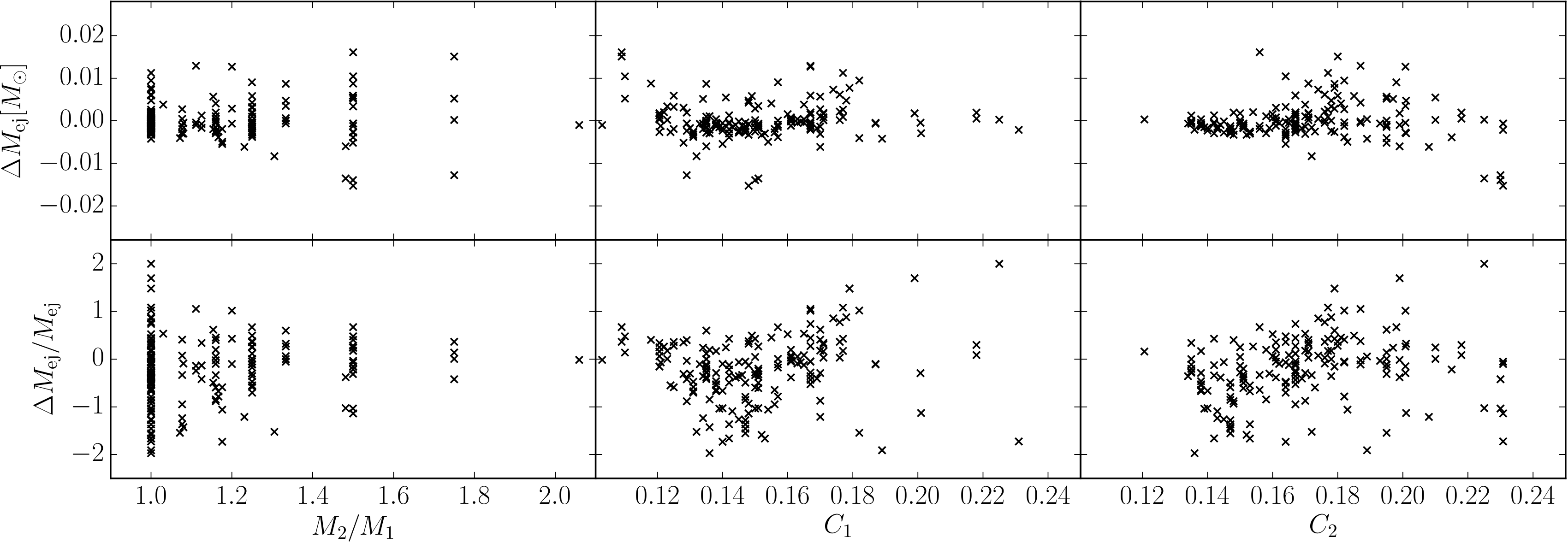}
  \caption{Difference between the ejecta mass of the NR simulation and the proposed fit. 
           Top panels show the absolute difference 
           $\Delta M_{\rm ej} = M_{\rm ej}^{\rm NR} - M_{\rm ej}^{\rm fit}$ 
           between the fit and the NR data and 
           bottom panels the relative difference 
           $2 \Delta M_{\rm ej}/(M_{\rm ej}^{\rm NR} + M_{\rm ej}^{\rm fit}$ . }
  \label{fig:fitMej}
  \end{center}
\end{figure}

Additionally, we present the results obtained from the fit 
in Fig.~\ref{fig:fitMej}, where the absolute and relative difference between the NR data 
and the fit are shown as a function of the mass ratio and the compactnesses of the stars. 
Obviously for equal mass setups the relative difference is larger because of the 
smaller ejecta mass. Those setups also have the highest NR uncertainty. 
Considering the influence of the compactnesses, we find that for 
larger compactness of the lighter star 
the absolute error increases. 

Let us also mention the possibility of obtaining fits for the ejecta mass 
(and other quantities) which are independent of the compactness of 
the stars and solely depend on the mass and tidal deformability, 
i.e.~on quantities directly accessible by a GW observation without assuming an EOS. 
One possibility might be the usage of quasi-universal compactness-Love
relations as mentioned in~\cite{Yagi:2016bkt} to substitute the 
compactness in \eqref{eq:Mej_fit}, also the baryonic mass
could be represented by the gravitational mass with introducing 
deviations to the NR only slightly larger than those of the current 
fits~\footnote{We thank Nathan~K.~Johnson-McDaniel for pointing this out.}. 
We are not following this approach here, since it did not allowed a better 
representation of the NR data and we tend to stay closer to the work previously 
presented for BHNSs systems. 

\subsection{Kinetic energy}

To estimate the kinetic energy of the ejecta we use a similar approach as for the 
unbound mass, i.e., 
\begin{equation}
 \frac{T_{\rm ej}^{\rm fit} } { 10^{50} {\rm erg}} =  \left[ a \left(\frac{M_2}{M_1}\right)^{1/3} \left(\frac{1-2 C_1}{C_1}\right)+ 
                          b \left(\frac{M_2}{M_1}\right)^{n}  + c \left(1 - \frac{M_1}{M^*_{1}}\right) 
                          \right] M_1^* + (1\leftrightarrow 2) + d . \label{eq:Tej_fit}
\end{equation}
The fitting parameters for the kinetic energy are: 
\begin{equation}
 a = -1.94315, \quad b = 14.9847,\quad c = -82.0025, \quad d = 4.75062, \quad n = -0.87914. \label{eq:par:Tej}
\end{equation}
The average residual between our fit and the pure NR data is 
$\Delta \bar{T}_{\rm ej} = 1.74\times10^{50}$erg, 
which corresponds to a difference of 79\%. 
Thus, the kinetic energy is slightly worse 
represented by our fit than the ejecta mass. 
The middle panels of figure~\ref{fig:Mej} represent our 
results for the kinetic energy, where again the 
1$\sigma$ and $2\sigma$ intervals are included 
($\Delta T_{\rm ej}^{1\sigma} = 2.4 \times 10^{50}$erg ).

\subsection{Ejecta velocities}

For the velocity we simplify our fitting function and restrict our analysis to the first 
66 data points in table~\ref{tab:overview}. For these data points the velocities inside the orbital plane 
and perpendicular to it are given. For BHNSs it is known that 
the velocity depends linearly on the mass ratio of the system, see~\cite{Kawaguchi:2016ana}. 
It was shown in~\cite{Dietrich:2016hky} that the same functional dependence holds for BNSs with 
high mass ratio or systems employing a stiff EOS. However, shock produced ejecta have a 
higher velocity component orthogonal to the orbital plane 
and should be included for a reliable estimate. 
Thus, we introduce an EOS dependent fitting function 
by including a first order polynomial depending on the compactness 
$(1+ c \ C_{1,2})$,  which leads to 
\begin{align}
 v_\rho & = \left[ a \left(\frac{M_1}{M_2}\right) \left(1 + c\  C_1\right)  \right] + (1\leftrightarrow 2) + b . \label{eq:vrho_fit}
\end{align}
The parameters are: 
\begin{equation}
 a = -0.219479, \quad b  = 0.444836, \quad c = -2.67385. \label{eq:par:vrho}
\end{equation}
Employing these parameters the NR data are represented with an average 
error of $\Delta \bar{v}_\rho = 0.020$, which corresponds to a
percentile difference of $13\%$. 

The same expression is used for the velocity orthogonal to the orbital plane:
\begin{align}
 v_z & = \left[ a \left(\frac{M_1}{M_2}\right) \left(1 + c\  C_1\right)  \right] + (1\leftrightarrow 2) + b . \label{eq:vz_fit}
\end{align}
As discussed, e.g.,~\cite{Hotokezaka:2012ze},
torque produced ejecta have much smaller velocities perpendicular 
to the orbital plane than inside the orbital plane. 
Thus, mostly shock driven ejecta cause large velocities orthogonal to the orbital plane. 
The parameters we obtain for $v_z$ are:  
\begin{equation}
 a = -0.315585, \quad b = 0.63808, \quad c = -1.00757\label{eq:par:vz}
\end{equation}
with average residuals of $\Delta v_z = 0.013$ and a fractional difference of $33\%$. 
The fractional difference is larger than for $v_\rho$ since the 
absolute value of the velocities is smaller. 

From $v_{\rho}$ and $v_{z}$ we estimate the total ejecta velocity as 
\begin{equation}
v_{\rm ej} = \sqrt{v_\rho^2+v_z^2}. \label{eq:vej}
\end{equation}
To check our description of $v_{\rm ej}$ we compare all data points 
(including the remaining 105 data points for which only 
the total ejecta velocity $v_{\rm ej}$ is known) to our fits. 
In total we obtain average residuals of 
$\Delta \bar{v}_{\rm ej} = 0.036$ 
and an average percentile uncertainty of 15\%. 
Figure~\ref{fig:Mej} (right panels) shows the ejecta velocities. 
We find that the residuals are smaller for the 66 data points which we used to obtain  
the fits of $v_\rho,v_z$ than for the remaining 105 data points. 
Overall one sees that the phenomenological fit slightly underestimates the velocity. 

\subsection{Other quantities}

\subsubsection{Geometry:} 
\label{sec:geometry}
The geometry of the ejecta can be extracted from NR simulations 
by considering 3D volume data of the density, 
but those data are not accessible for most of the 
configurations presented in table~\ref{tab:overview}. 
Thus, we want to present in the following a model for homogeneously 
distributed material inside an annular sector moving with the velocity $v_{\rm ej}$. 
Inside the $\rho-z$-plane the ejecta is distributed in a circular sector 
with a polar opening angle $2 \theta_{\rm ej}$.
The ejected material has an azimuthal opening angle of $\phi_{\rm ej}$. 
Under the assumption that the ejecta consists of 
particles moving radially outward with velocity $v_{\rm ej}$, we obtain by averaging over 
all particles the following equations for $v_\rho$ and $v_z$:
\begin{equation}
 v_\rho  \approx  v_{\rm ej} \frac{\sin{(\theta_{\rm ej})}}{\theta_{\rm ej}} , \quad  v_z    \approx v_{\rm ej} \frac{1-\cos{(\theta_{\rm ej})}}{\theta_{\rm ej}}. \label{eq:theta1}
\end{equation}
For a non-zero, but small $\theta_{\rm ej}$ one gets
\begin{equation}
 \frac {\theta_{\rm ej}^3}{24} +  \frac{\theta_{\rm ej}}{2} - \frac{v_z}{v_\rho} \approx 0, 
\end{equation}
which can be solved for $\theta_{\rm ej}$:
\begin{equation}
\theta_{\rm ej} \approx \frac{ - 2^{4/3} v_\rho^2 + 2^{2/3} 
(v_\rho^2 ( 3 v_z + \sqrt{ 9 v_z^2 + 4 v_\rho^2}))^{2/3} }
{(v_\rho^5(3 v_z + \sqrt{9 v_z^2 + 4v_\rho^2}))^{1/3}}.  \label{eq:theta_fit}
\end{equation}

\begin{figure}[t]
  \begin{center}
  \includegraphics[width=0.95\textwidth]{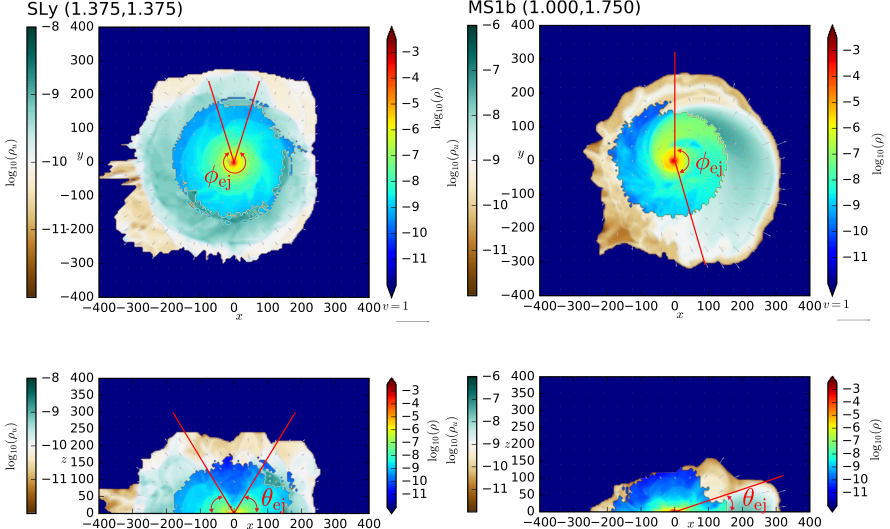} 
  \caption{2D density plots with rest mass $\rho$ shown from blue to 
           red with increasing density and 
           the unbound material $\rho_u$ shown brown to green with increasing density. 
           Geometric units are employed. 
           We use the velocity as extracted from the numerical simulation and show 
           $\theta_{\rm ej}$ and $\phi_{\rm ej}$ as approximated from \eqref{eq:theta_fit} and 
           \eqref{eq:phi_fit}. 
           Left: Simulations \#66 
           (SLy,1.375$M_\odot$,1.375$M_\odot$)
           Right: Simulation \#55 (MS1b,1.000$M_\odot$,1.750$M_\odot$). }
  \label{fig:geometry}
  \end{center}
\end{figure}

In contrast to the opening angle $\theta_{\rm ej}$, it is more difficult 
from our current results to estimate the azimuthal 
angle $\phi_{\rm ej}$. In~\cite{Kawaguchi:2016ana} was assumed that 
BHNS setups have an azimuthal angle of  $\phi_{\rm ej} \approx \pi$. 
This is in agreement with high mass ratio BNS mergers employing stiff EOSs~\cite{Dietrich:2016hky}, 
i.e.~for setups where torque is the dominant ejection mechanism. 
Contrary if shock ejecta are present, e.g.~for softer EOSs, 
the azimuthal angle even increases up to $2 \pi$, 
i.e.~there exists a correlation between $\theta_{\rm ej}$ and $\phi_{\rm ej}$. 
Assuming that the opening angles vary between 
$\theta_{\rm ej} \in [\pi/8,3 \pi/8]$ and 
$\phi_{\rm ej} \in [\pi,2\pi]$, and that $\theta_{\rm ej}$ and $\phi_{\rm ej}$ are linearly correlated, we obtain
\begin{equation}
 \phi_{\rm ej} = 4 \theta_{\rm ej} + \frac{\pi}{2}.  \label{eq:phi_fit}
\end{equation}

To test our approximations, we present snapshots of the density profile 
in the $x$-$y$ and $x$-$z$ plane for the simulations \#55 and \#66 in figure~\ref{fig:geometry}. 
We show the rest-mass density $\rho$ (color bar ranging from blue to red) 
and the unbound rest mass density $\rho_u$ (color bar ranging from brown to green). 
The two cases present two rather extreme setups, namely a stiff EOS 
with a large mass ratio and a soft EOS for an equal mass system. 
In figure~\ref{fig:geometry} we also include the approximations for $\theta_{\rm ej}$ 
and $\phi_{\rm ej}$ obtained from \eqref{eq:theta_fit} and \eqref{eq:phi_fit}. 
The examples show that the geometry of the higher density ejecta regions can be
described reasonably well with our model.

\subsubsection{Composition:}
\label{sec:composition}
Caused by different ejecta mechanisms the composition and electron fraction of the ejecta 
varies depending on the EOS, mass ratio, and total mass. 
As pointed out in the literature, unbound material ejected due to torque in the tidal tail of the NSs has 
a low electron fraction, see e.g.,~\cite{Rosswog:2015nja}. 
Contrary ejecta produced via shock heating have overall 
a broader range in electron fraction, e.g.,~\cite{Sekiguchi:2016bjd}.
Table~\ref{tab:Ye} shows the fraction of data from table~\ref{tab:overview} for which we also know the 
average electron fraction. 
Note that the electron fraction of the ejected material varies significantly
among different implementations for the neutrino transport, 
e.g.,~\cite{Radice:2016dwd,Palenzuela:2015dqa,Lehner:2016lxy} 
find overall smaller electron fractions of the unbound material 
than reported in~\cite{Sekiguchi:2016bjd}. 
Consequently the presented results have to be taken with care and 
the following should be regarded as a qualitative
discussion. 

  \begin{table}[t]
    \caption{ \label{tab:Ye} Columns refer to: 
    The data ID as in table~\ref{tab:overview}, 
    the mass of the first star $M_1$, 
    the mass of the second star $M_2$, 
    the ejecta mass $M_{\rm ej}$, 
    the kinetic energy of ejecta $T_{\rm ej}$, 
    the ejecta velocity $v_{\rm ej}$, and 
    the electron fraction $Y_e$. 
    All setups have been simulated in~\cite{Sekiguchi:2016bjd}.}
    \begin{center}
\begin{small}
    \begin{tabular}{l|ccc|cccc}  
    $\#$ & EOS & $M_1$     & $M_2$     &  $M_{\rm ej}$       & $T_{\rm ej}$  & $v_{\rm ej}$ & $Y_e$\\ 
         &     &$[M_\odot]$&$[M_\odot$]& $[10^{-3}M_\odot]$  & $[10^{50}$erg]& $[c]$ & \\  
    \hline
85  & DD2   &  1.25  &  1.45  &  5    &   1.61  &  0.19  &  0.2  \\
88  & DD2   &  1.3   &  1.4   &  3    &   0.87  &  0.18  &  0.26 \\
98  & DD2   &  1.35  &  1.35  &  2    &   0.46  &  0.16  &  0.3  \\
145 & SFHo  &  1.25  &  1.45  &  11   &   5.66  &  0.24  &  0.18 \\
148 & SFHo  &  1.3   &  1.4   &  6    &   2.15  &  0.2   &  0.27 \\
149 & SFHo  &  1.33  &  1.37  &  9    &   3.55  &  0.21  &  0.3 \\
158 & SFHo  &  1.35  &  1.35  &  11   &   4.76  &  0.22  &  0.31 \\
  \end{tabular}
\end{small}
  \end{center}
\end{table}

\begin{figure}[t]
  \begin{center}
  \includegraphics[width=1\textwidth]{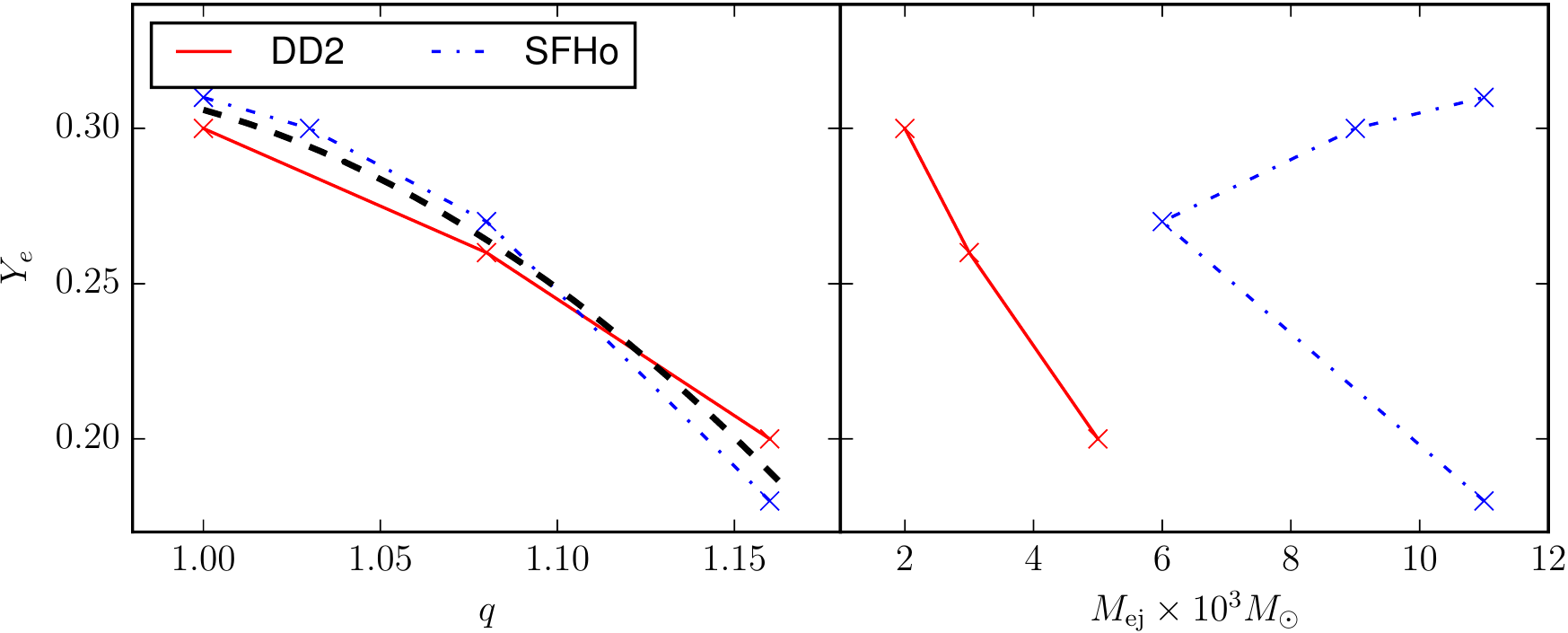}
  \caption{Left panel: Electron fraction $Y_e$ as a function of the mass ratio $q$. 
           Right panel: Electron fraction $Y_e$ as a function of the ejecta mass $M_{\rm ej}$. 
           We present data for two different EOSs: SFHo (blue dashed dotted line ) and 
           the stiffer DD2 (red solid line).
           In the left panel we also include as a black dashed line the fit of \eqref{eq:fit:Ye}.}
  \label{fig:Ye}
  \end{center}
\end{figure}

Figure~\ref{fig:Ye} summarized the important results from table~\ref{tab:Ye}. 
As shown in figure~\ref{fig:EOS} the DD2 EOS is softer than SFHo. 
Considering the left panel of figure~\ref{fig:Ye} we observe that for both EOSs 
an increasing mass ratio leads to a smaller electron fraction. 
This is expected since more ejecta are produced due to torque independent of the EOS. 
The right panel shows the dependence between the ejecta mass and the electron fraction. 
For all setups more massive ejecta are produced for the softer EOS, e.g., for $q=1$ more 
than five times more mass is ejected for the SFHo EOS. 
For this mass ratio the dominant ejection mechanism for SFHo 
is shock heating, which seems to be suppressed for increasing 
mass ratios. Thus, the ejecta mass and the electron fraction decreases 
for increasing $q$ (see also the explanation in~\cite{Sekiguchi:2016bjd}). 
Interestingly is that while for DD2 $Y_e(M_{\rm ej})$ is monotonic, this is not true for 
SFHo, where beyond a mass ratio of $q\approx1.1$ the ejecta mass is growing again. 
We propose that for $q> 1.1$ also SFHo setups become dominated by torque produced ejecta 
and shocks are suppressed. 

Finalizing our consideration of the composition, 
we want to present a fit for the electron fraction as a function of the 
mass ratio for a total mass of $M=2.7M_\odot$ for the data of~\cite{Sekiguchi:2016bjd}:
\begin{equation}
 Y_{e} = 0.306 - 0.318 (q-1) - 2.568 (q-1)^2  \label{eq:fit:Ye}.
\end{equation}
The fit is shown as a black dashed line in figure~\ref{fig:Ye} (left panel).
To generalize~\eqref{eq:fit:Ye} to different 
total masses and higher mass ratios more simulations
including realistic microphysical treatments are required. 

\subsubsection{Spin effects:}
\label{sec:spin}

\begin{figure}[t]
  \begin{center}
  \includegraphics[width=1\textwidth]{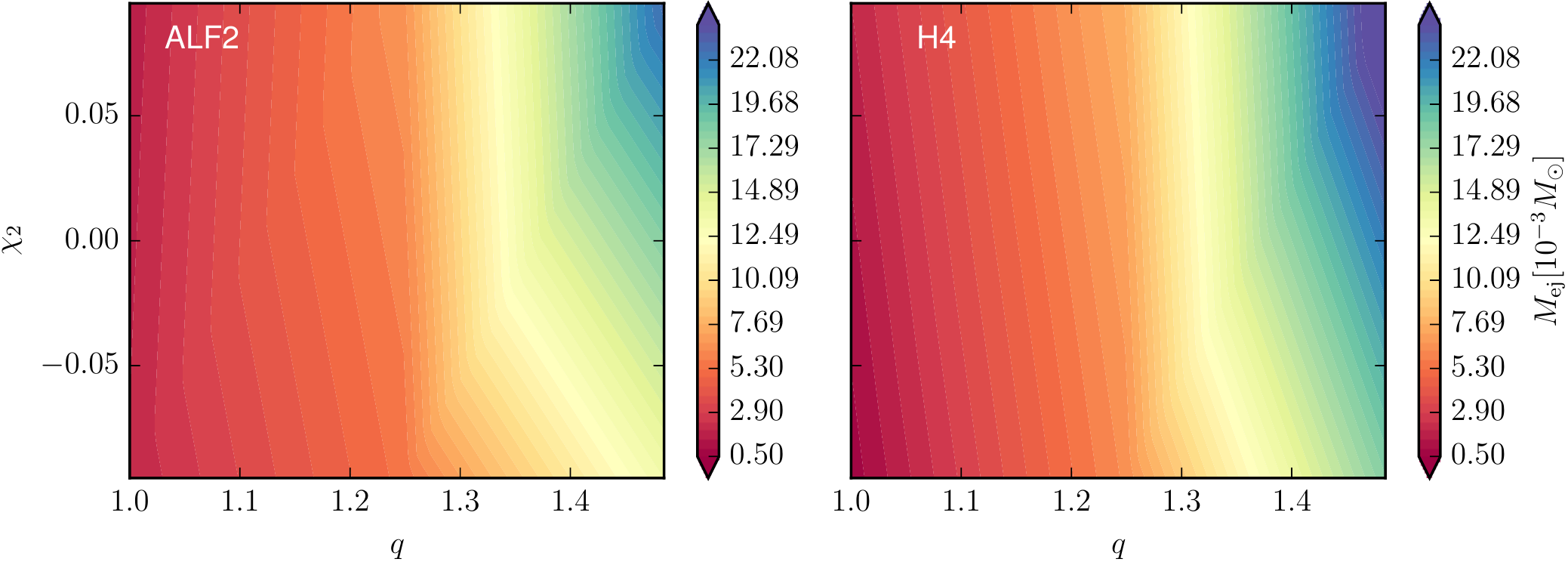}
  \caption{Ejecta mass for the spinning configurations of table~\ref{tab:Spin} 
           as a function of the mass ratio $q$ and the spin of 
           the secondary star $\chi_2$ for the ALF2 EOS (left) 
           and the H4 EOS (right).}
  \label{fig:Spin}
  \end{center}
\end{figure}

Let us also briefly comment on the effect of the star's intrinsic rotation on the 
ejecta quantities. We summarize in tab.~\ref{tab:Spin} the spinning configurations 
of~\cite{Dietrich:2016lyp}. 
Figure~\ref{fig:Spin} visualizes these data and shows the influence of the mass ratio 
and of the spin of the secondary (less massive star) 
on the ejecta mass. 
The figure shows two distinct effects 
(i) for an increasing mass ratio more material becomes unbound (as already discussed above), 
(ii) if the spin of the secondary star is aligned to the orbital angular momentum
(positive) then the ejecta mass increases even further. 

As pointed out in~\cite{Dietrich:2016lyp} 
spin aligned to the orbital angular momentum enhances the ejection, 
while contrary antialigned spin leads to lower massive ejecta. 
This can be understood by considering the fluid velocity inside the tidal tail,
which at lowest order can be approximated as the sum of the orbital fluid velocity 
and the fluid velocity connected to the intrinsic rotation of the star.
In cases where the individual star also has spin parallel to the orbital 
angular momentum the fluid velocity inside the tail is higher and consequently material 
gets unbound and leaves the system. 
This effect becomes most prominent for systems for which material ejection is caused 
by torque, e.g.~by unequal mass systems. 
Because in unequal mass systems the mass ejection happens mostly from 
the tidal tail of the lower massive star,
the determining quantity is the spin of the secondary star $\chi_2$
as shown in figure~\ref{fig:Spin}.

  \begin{table}[t]
  \caption{ \label{tab:Spin} Overview about the spinning simulations taken from~\cite{Dietrich:2016lyp}.
  The columns refer to: EOS, individual masses $M_{1,2}$, dimensionless spins of the stars $\chi_{1,2}$, 
  the ejecta mass $M_{\rm ej}$, kinetic energy of the ejecta $T_{\rm ej}$, 
  velocity inside the orbital plane $v_\rho$ and perpendicular to it $v_z$.}
\begin{small}
\begin{center}
  \begin{tabular}{ccccc|cccc} 
    EOS & $M_1$       & $\chi_1$   & $M_2$       & $\chi_2$ & $M_{\rm ej}$        & $T_{\rm ej}$  & $v_{\rho}$ & $v_{z}$ \\ 
        & $[M_\odot]$ &            &  $[M_\odot$]&                          & $[10^{-3}M_\odot]$  & $[10^{50}$erg]& $[c]$      & $[c]$   \\  
     \hline
ALF2 & 1.375 & 0.102 & 1.375 &  -0.102 &  4.1 & 0.55 &  0.12 & 0.07  \\
ALF2 & 1.375 & 0.102 & 1.375 &   0.000 & 2.0 & 0.36 &  0.13 & 0.05  \\
ALF2 & 1.375 & 0.102 & 1.375 &   0.102 & 1.6 & 0.32 &  0.16 & 0.05  \\
ALF2 & 1.528 & 0.104 & 1.223 &  -0.102 & 4.5 & 1.7  &  0.15 & 0.11  \\
ALF2 & 1.528 & 0.104 & 1.222 &  0.000  & 5.5 & 2.1  &  0.16 & 0.13  \\
ALF2 & 1.528 & 0.104 & 1.223 &  0.102  & 6.7 & 2.   &  0.16 & 0.08  \\
ALF2 & 1.651 & 0.107 & 1.100 &  -0.101 & 11  & 3.6  &  0.18 & 0.05   \\
ALF2 & 1.651 & 0.107 & 1.100 &   0.000 & 14  & 4.1  &  0.18 & 0.04   \\
ALF2 & 1.651 & 0.107 & 1.100 &   0.101 & 24  & 7.5  &  0.18 & 0.04   \\
H4   & 1.375 & 0.100 & 1.375 &  -0.100 & 1.5 & 0.62 &  0.16 & 0.10  \\
H4   & 1.375 & 0.100 & 1.375 &  0.000  & 0.7 & 0.23 &  0.17 & 0.10  \\
H4   & 1.375 & 0.100 & 1.375 &  0.100  & 2.0 & 0.78 &  0.15 & 0.07  \\
H4   & 1.528 & 0.100 & 1.223 &  -0.100 & 4.1 & 1.7  &  0.17 & 0.09  \\
H4   & 1.528 & 0.100 & 1.222 &  0.000  & 6.4 & 3.2  &  0.18 & 0.08  \\
H4   & 1.528 & 0.100 & 1.223 &  0.100  & 7.8 & 3.0  &  0.18 & 0.11  \\
H4   & 1.651 & 0.101 & 1.100 &  -0.099 & 9.5 & 2.4  &  0.17 & 0.03   \\
H4   & 1.651 & 0.101 & 1.100 &   0.000 & 19  & 5.5  &  0.17 & 0.03   \\
H4   & 1.651 & 0.101 & 1.100 &   0.099 & 27  & 7.5  &  0.17 & 0.02   \\
 \end{tabular}
 \end{center}
\end{small}
 \end{table}

\section{Kilonovae}
\label{sec:kilonovae}

It is expected that the ejected material is heated up 
because of the radioactive decay of r-process elements 
and consequently triggers EM emission called kilo- or macronovae, see among 
others~\cite{Li:1998bw,Metzger:2010sy,Roberts:2011xz,Goriely:2011vg,Tanvir:2013pia,Korobkin:2012uy,Grossman:2013lqa,Tanaka:2013ana,Tanaka:2016sbx,Rosswog:2016dhy}
and for overview articles~\cite{Fernandez:2015use,Metzger:2016pju}.
Up to date there are three possible kilonovae candidates
for which a connection to a GRB has been made: 
GRB 050709~\cite{Jin:2016pnm}, GRB 060614~\cite{Yang:2015pha}, 
GRB 130603B~\cite{Tanvir:2013pia}. 
The most likely origin of these kilonovae candidates 
are compact binary mergers. 

\subsection{Peak quantities}

\begin{figure}[t]
  \includegraphics[width=\textwidth]{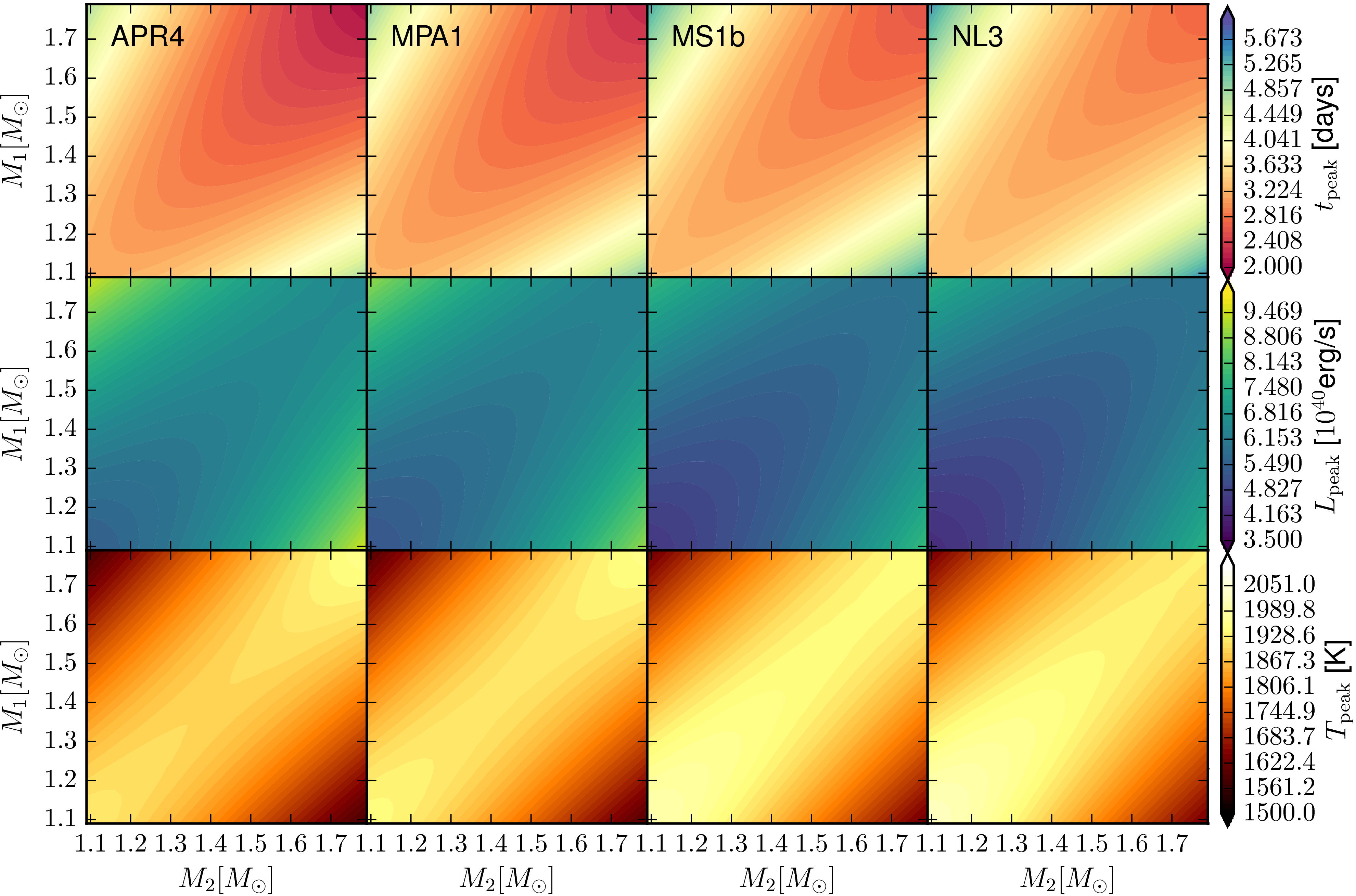}
  \caption{Kilonovae properties: upper panel shows the time when the 
  peak luminosity is reached; middle panels show the corresponding luminosity, 
  and the bottom panel the corresponding temperature. 
  We present results for four different EOSs, from left to right: 
  APR4, MPA1, MS1b, NL3, i.e., the compactness is from left to right decreasing, 
  see figure~\ref{fig:EOS}. The quantities are given in terms of the 
  individual masses of the stars $M_1,M_2$.}
  \label{fig:macro}
\end{figure}

Based on the work of~\cite{Grossman:2013lqa} we will present some important kilonovae properties. 
The time $t_{\rm peak}$ at which the peak in the near-infrared occurs,
the bolometric luminosity at this time $L_{\rm peak}$, and the corresponding temperature
$T_{\rm peak}$ are given as:
\numparts \begin{align}
   t_{\rm peak} & = 4.9 \ {\rm days}  \times \left( \frac{M_{ej}}{10^{-2} M_\odot} \right)^{\frac{1}{2}} 
                      \left( \frac{\kappa}{10 {\rm cm^2 g^{-1}} } \right)^{\frac{1}{2}}
                      \left( \frac{v_{\rm ej}}{0.1} \right)^{-\frac{1}{2}} , \label{eq:tpeak} \\
   L_{\rm peak}&  = 2.5 \cdot  10^{40} {\rm erg \, s^{-1}} \times       \left( \frac{M_{ej}}{10^{-2} M_\odot} \right)^{1-\frac{\alpha}{2}} 
                      \left( \frac{\kappa}{10 {\rm cm^2 g^{-1}} } \right)^{-\frac{\alpha}{2}}
                      \left( \frac{v_{\rm ej}}{0.1} \right)^{\frac{\alpha}{2}} , \label{eq:Lpeak} \\
   T_{\rm peak} & = 2200 {\rm K}    \times \left( \frac{M_{ej}}{10^{-2} M_\odot} \right)^{-\frac{\alpha}{8}} 
                      \left( \frac{\kappa}{10 {\rm cm^2 g^{-1}} } \right)^{-\frac{\alpha+2}{8}}
                      \left( \frac{v_{\rm ej}}{0.1} \right)^{\frac{\alpha-2}{8}} . \label{eq:Kpeak} 
\end{align} \endnumparts
In \cite{Grossman:2013lqa} the authors assume that the energy release due to
the radioactive decay is proportional to $\sim t^{-\alpha}$ with
$\alpha=1.3$. We set the average opacity to $\kappa =
10~{\rm cm^2 g^{-1}}$~\footnote{Notice that 
as shown in e.g.,~\cite{Tanaka:2013ana,Kasen:2013xka}
the typical opacity for a kilonovae is significantly higher than for 
typical supernovae explosions, which is caused by the 
presence of lanthanides.
The exact value of the opacity depends on the composition of the 
material, which is not included in our models.}.

In figure~\ref{fig:macro} we present $t_{\rm peak},L_{\rm peak},T_{\rm peak}$ 
for four different EOSs as a function of the individual masses $M_1,M_2$. 
We find for all setups that an increasing mass-ratio increases $t_{\rm peak}$, $L_{\rm peak}$ and 
decreases $T_{\rm peak}$. 
Furthermore an increasing total mass leads to a decreasing 
$t_{\rm peak}$. 
Considering the influence of the EOS, softer EOSs lead to more luminous kilonovae 
in particular for equal mass merger. This can be explained by smaller ejecta mass 
caused by the absence of shock driven ejecta for stiff EOSs. For systems close to 
equal mass the temperature of the kilonovae is higher. 
Interesting is also that for equal mass systems the luminosity and the temperature have 
saddle points, see middle and lower panels. 
This means that keeping the mass ratio fixed 
a local extrema exist for which the luminosity becomes maximal 
and that also a local extrema exists for 
which the temperature becomes minimal. Both points do not have to coincide.
It would be interesting to test with further NR simulations whether such a saddle 
point exists or is just an artifact of the employed fit. 

\subsection{Time evolution}

\subsubsection{Luminosity:}

\begin{figure}[t]
  \begin{center}
  \includegraphics[width=0.8\textwidth]{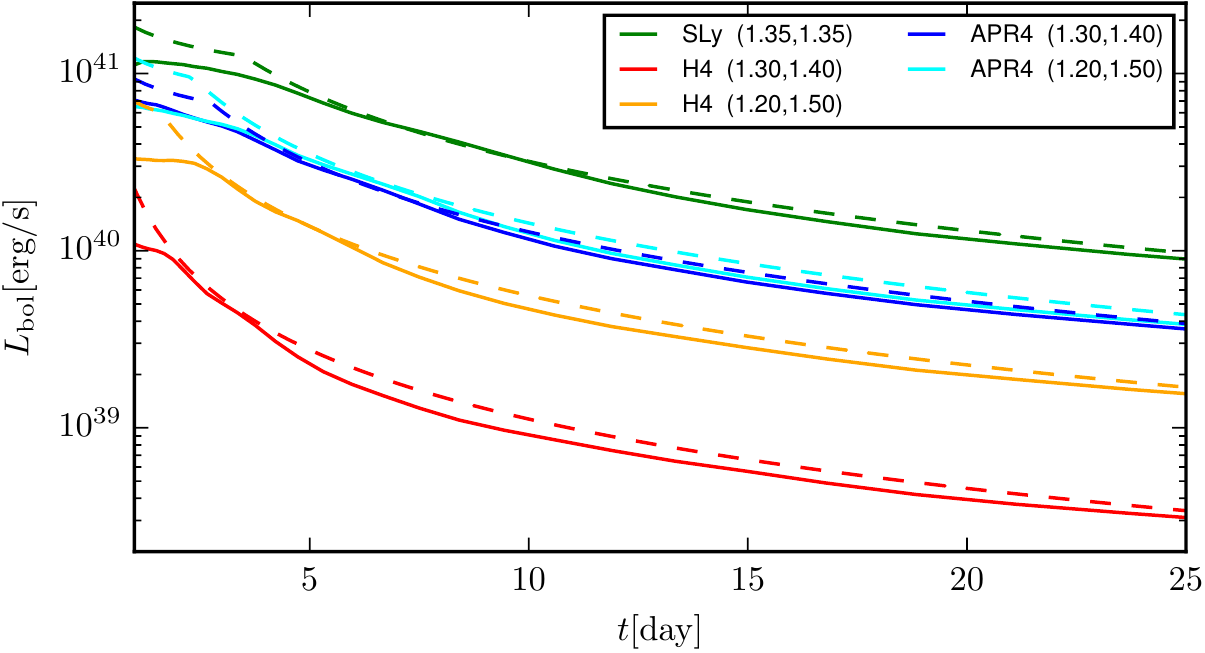}
  \caption{Comparison of the bolometric luminosity given by \eqref{eq:Lbol} (dashed lines)
  and a radiative transfer simulation (solid lines). The results of the radiative transfer simulation 
  was presented in~\cite{Tanaka:2013ana,Tanaka:2013ixa} and is public available at~\cite{Tanaka_web}.
  The legend characterizes the EOS and the individual masses of the NSs are given in solar masses.}
  \label{fig:Lbol}
  \end{center}
\end{figure}

To determine the luminosity of the kilonovae, we follow the discussion 
of~\cite{Kawaguchi:2016ana}, which we briefly summarize below. 
As described in section~\ref{sec:geometry} the ejecta is modeled as a 
partial sphere in the latitudinal and longitudinal direction.
We further assume that the material is homogeneously distributed inside the ejecta and that 
photons purely escape from the latitudinal edge. This agrees with the assumptions 
made in~\cite{Kawaguchi:2016ana} and also gives valid results for BNS mergers
as shown below. 
Considering that the optical depth increases with decreasing density, 
the whole region becomes visible after  
\begin{equation}
 t_c = \sqrt{\frac{\theta_{\rm ej} \kappa M_{\rm ej}}{2 \phi_{\rm ej} (v_{\rm max}-v_{\rm min})}},
\end{equation}
with $v_{\rm max}$, $v_{\rm min}$ being the maximum 
and the minimum speed of the ejecta. 
The mass of the photon escaping region is then given by 
$M_{\rm obs} = M_{\rm ej} (t/t_c)$ for times $t<t_c$. 
In \cite{Korobkin:2012uy,Wanajo:2014wha} 
was shown that the specific heating for energy release caused by 
radioactive decay can be approximated by 
$
 \dot{\epsilon} \approx \dot{\epsilon}_0 \left(\frac{t}{\rm 1 day}\right)^{-\alpha}.
$
This allows to write the bolometric luminosity as 
\begin{equation}
 L(t) = (1 + \theta_{\rm ej}) \epsilon_{\rm th} \dot{\epsilon}_0 M_{\rm ej} 
              \begin{cases}
               \frac{t}{t_c} \left(\frac{t}{1 \ {\rm day}}\right)^{-\alpha}, \quad t \leq t_c \\
               \left(\frac{t}{1\ \rm day}\right)^{-\alpha}, \quad t > t_c \\
            \end{cases},  \label{eq:Lbol}
\end{equation}
where we will use $\dot{\epsilon}_0=1.58 \times 10^{10} {\rm erg \ g^{-1}\ s^{-1}}$ and $\alpha=1.3$
for our considerations~\footnote{Note that as discussed in~\cite{Kawaguchi:2016ana} \eqref{eq:Lbol}
also used the assumption of a small opening angle $\theta_{\rm ej}$ which is valid for BHNSs 
but might be violated for BNS systems. However, figure~\ref{fig:Lbol} 
reveals that reasonable results are also obtained for BNS systems with larger opening angles, 
see e.g., SLy (1.35,1.35).}.

In figure~\ref{fig:Lbol} a comparison between \eqref{eq:Lbol} 
and the radiative transfer simulations of~\cite{Tanaka:2013ana,Tanaka_web}
is presented. One sees remarkable agreement between the simple 
model function and the radiative transfer simulations. 
As input variables for \eqref{eq:Lbol}, 
we have used the stated ejecta masses from~\cite{Tanaka_web}. 
This is necessary since $L_{\rm bol}$ depends strongly on $M_{\rm ej}$ 
such that a difference in $M_{\rm ej}$ produces a large difference in 
$L_{\rm bol}$ and a comparison would not test the assumptions made for \eqref{eq:Lbol}, 
but how \eqref{eq:Mej_fit}
approximates this particular setup. 
Furthermore, $v_{\rm min}$ is set to 0.02, 
$v_{\rm max} = 2 v_{\rm ej} - v_{\rm min}$, 
and $\theta_{\rm ej}$ and $\phi_{\rm ej}$ are 
chosen according to \eqref{eq:theta_fit} and \eqref{eq:phi_fit}. 
Figure~\ref{fig:Lbol} proves that \eqref{eq:Lbol}, which was 
originally proposed for BHNS setups in~\cite{Kawaguchi:2016ana}
also allows to describe BNS mergers and the time evolution of the kilonovae. 

\subsubsection{Lightcurves:}

From the given luminosity the bolometric magnitude can be computed according to: 
\begin{equation}
 M_{\rm bol} \approx 4.74 -2.5 \log_{10} \left( \frac{L_{\rm bol}}{L_\odot}\right), \label{eq:Mbol}
\end{equation}
with $L_\odot$ denoting the bolometric luminosity of the sun. 
To compute the magnitude in each wavelength, 
we have to know the spectra of the kilonovae. 
One possible approach to compute the spectra is by considering 
the effective temperature of the photosphere 
\begin{equation}
 T \approx \left( \frac{L(t)}{\sigma S(t)}\right)^{1/4},
\end{equation}
with $S(t)$ being the surface of the latitudal edge, and to assume 
that the spectrum of a kilonovae can be approximated by 
a pseudo black body spectrum, e.g.,~\cite{Kasen:2013xka}. 

\begin{figure}[t]
  \begin{center}
  \includegraphics[width=1\textwidth]{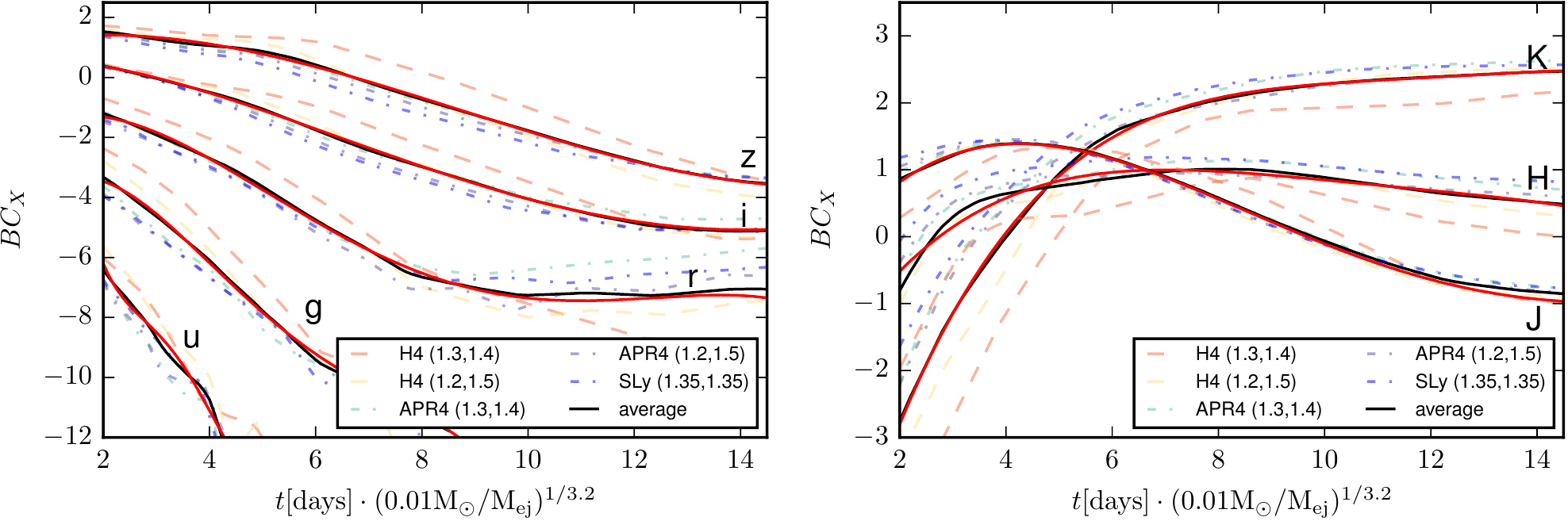}
  \caption{Bolometric corrections for the ugriz-bands (left) and 
           KHJ-bands (right) as a function of the rescaled time 
           $t'=t [{\rm days}](0.01M_\odot/M_{\rm})^{1/3.2}$.
           We use public available results of~\cite{Tanaka_web}
           and show them as dashed and dot-dashed lines. 
           The average of the available data for each individual band is shown 
           as a black solid line and a fit of the average is visible as a
           red solid line. The parameters for the fit are given in
           \eqref{eq:BC_z}-\eqref{eq:BC_J}.}
  \label{fig:BC}
  \end{center}
\end{figure}

Another approach enabling us to compute the spectrum 
are bolometric corrections (BC) as discussed in~\cite{Kawaguchi:2016ana}.
The final magnitude in each band (denoted by the subscript $X$) is then given by
\begin{equation}
 M_X(t) = M_{\rm bol}(L(t)) - BC_X(t).
\end{equation}
To compute the bolometric corrections we use the 
public available light curves of~\cite{Tanaka_web}. 
It was shown in~\cite{Kawaguchi:2016ana}
that the time evolution of the BCs for BHNSs agrees once 
the elapsed time is rescaled by 
$t'= t \cdot (10^{-2} M_\odot/M_{\rm ej})^{1/3.2}$. 
Figure~\ref{fig:BC} shows that the same rescaling 
can be used for BNS data. 
We present for five different setups~\cite{Tanaka_web} the BCs for the ugriz-band 
in the left and for the KHJ-band in the right panel.
The difference among the different setups of the BC is about 1 magnitude. 
To obtain the final BC, we average the results of all five configurations 
(black solid line) and fit 
the average with a polynomial (red solid lines)
\begin{equation}
 BC_X = a_0 + a_1 t' + a_2 t'^2 + a_3 t'^3 + a_4 t'^4 .
\end{equation}
The final parameters for the polynomials fits are
\numparts \begin{eqnarray}
 BC_z: & (1.072,0.3646,-0.1032,0.00368,0.0000126)    & t' \in [2,15] \label{eq:BC_z}\\ 
 BC_i: & (0.6441,0.0796,-0.122,0.00793,-0.000122)     & t' \in [2,15]  \\  
 BC_r: & (-2.308,1.445,-0.5740,0.0531,-0.00152)      & t' \in [2,15]  \\  
 BC_g: & (-6.195,4.054,-1.754,0.2246,-0.009813)     & t' \in [2,8.5]  \\ 
 BC_u: & (40.01,-56.79, 25.73,-5.207,0.3813)         & t' \in [2,5] \\    
 BC_K: & (-7.876,3.245, -0.3946,0.0216,-0.000443)    & t' \in [2,15] \\ 
 BC_H: & (-2.763,1.502,-0.2133,0.0128,-0.000288)     & t' \in [2,15] \\
 BC_J: & (-1.038,1.348,-0.2364,0.0137,-0.000261)     & t' \in [2,15]   \label{eq:BC_J}.
\end{eqnarray} \endnumparts

\begin{figure}[t]
  \begin{center}
  \includegraphics[width=1\textwidth]{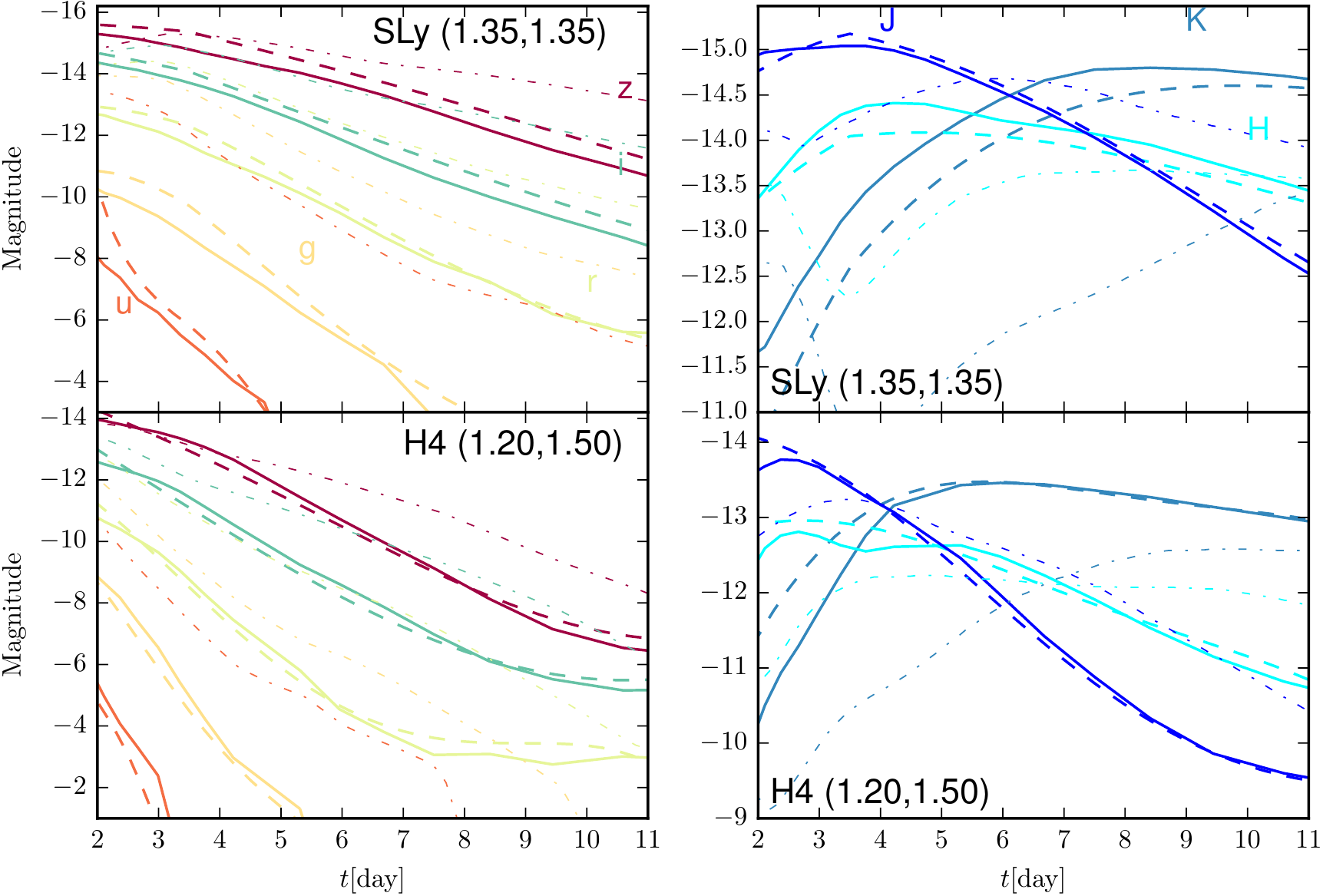}
  \caption{Absolute Magnitudes in the ugridz-bands (left panels) and JHK-bands (right panels)
           for the equal mass SLy (1.35,1.35) and the unequal mass H4 (1.20,1.50) setups. 
           The solid lines represent the data reported in~\cite{Tanaka:2013ana,Tanaka_web}. 
           The dashed lines represent data obtained from \eqref{eq:Lbol} including 
           the computed BC corrections. 
           We also include as a thin dashed dotted line results obtained with the 
           public available code of~\cite{Kawaguchi_web}.}
  \label{fig:lightcurve}
  \end{center}
\end{figure}

As an example we compare the lightcurves obtained from the discussed model and 
computed with the radiative MC code of~\cite{Tanaka:2013ana,Tanaka_web} 
for two systems: one equal mass system employing a soft EOS (SLy $(1.35M_\odot,1.35M_\odot$) ) 
and one unequal masses case with a stiffer EOS (H4 $(1.20M_\odot,1.50M_\odot)$ ). 
As for figure~\ref{fig:Lbol} we use here the ejecta mass stated in~\cite{Tanaka_web} to 
compute the bolometric luminosities.
Figure~\ref{fig:lightcurve} shows that after applying the BCs, 
the MC results and those obtained by the simple model agree well.
Additionally, we also include lightcurves computed with the public available code 
of~\cite{Kawaguchi_web} (thin dot dashed lines), which was developed for BHNS mergers 
and which shows a larger disagreement to the MC results. 
The difference between the MC simulation and the model presented here is smaller
because of the particular choice of the BCs. 

\section{Radio flares}
\label{sec:radio}

\begin{figure}[t]
  \includegraphics[width=\textwidth]{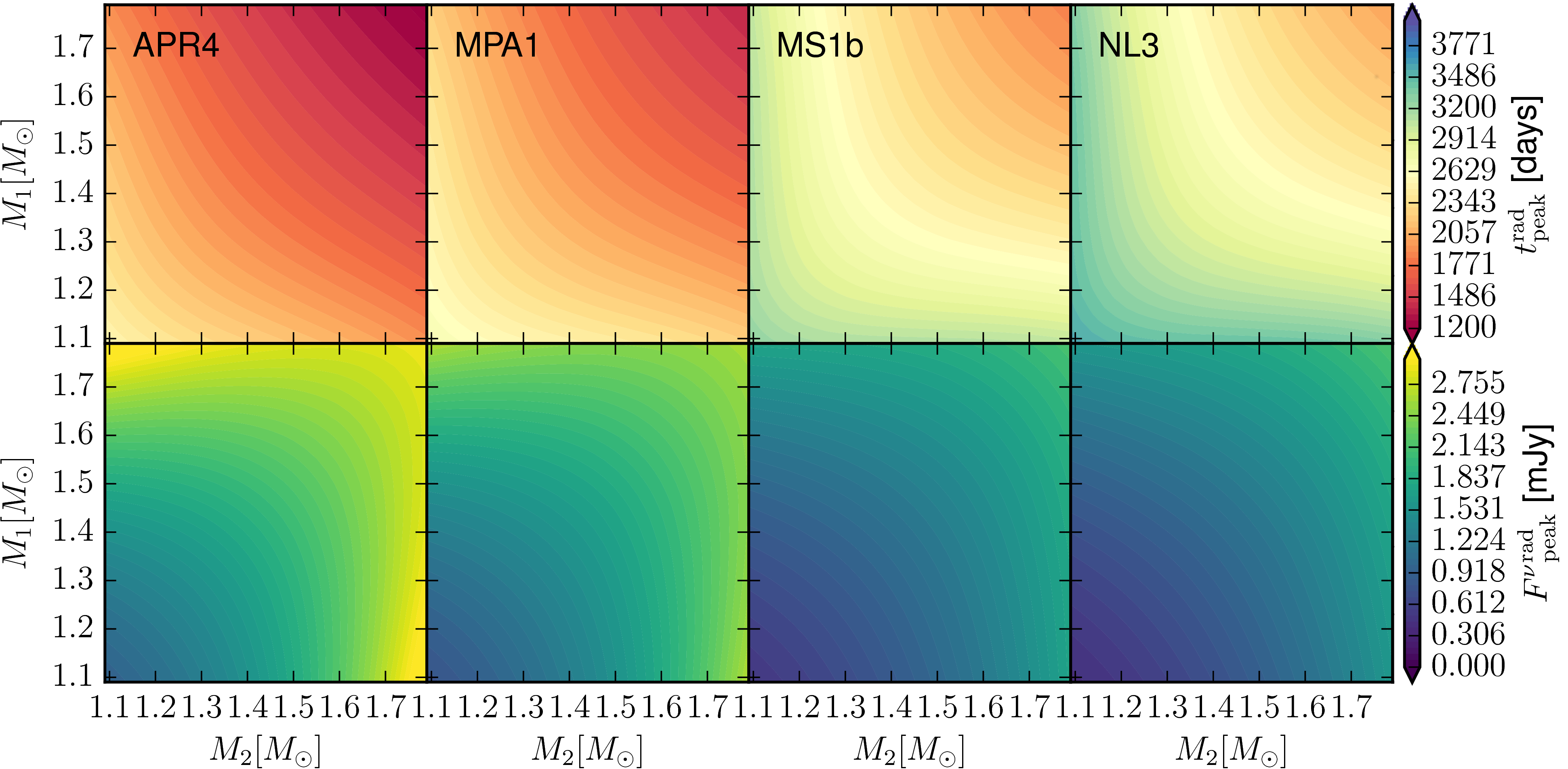}
  \caption{Radio flares properties: upper panel shows the time once the 
  peak in the radio band is observable after the merger of the two neutron stars; 
  lower panel shows the radio fluency at this time. 
  We present results for four different EOSs, from left to right: 
  APR4, MPA1, MS1b, NL3, i.e., the compactness is from left to right decreasing, 
  see figure~\ref{fig:EOS}. The quantities are given in terms of the 
  individual masses of the stars $M_1,M_2$.}
  \label{fig:radio}
\end{figure}

In addition to kilonovae, it is possible that
sub-relativistic outflows produce radio flares 
with peak times of a few month up to years 
after the merger of the compact binary. 

In order to estimate the radio emission, 
we use the model of~\cite{Nakar:2011cw}. 
The strongest signal is expected at a time 
\begin{align}
   t_{\rm peak}^{\rm rad} & = 1392 \ {\rm days} \times \left( \frac{T_{\rm ej} }{10^{49}{\rm erg}} \right)^{\frac{1}{3}} 
                  \left( \frac{n_0}{\rm cm^{-3}}\right)^{-\frac{1}{3}} 
                  \left( \frac{v_{\rm ej}}{0.1} \right)^{-\frac{5}{3}} 
\end{align}
after the merger of the system. 
The radio fluence at this time is 
\begin{align}
   {F^\nu}^{\rm rad}_{\rm peak} & = 0.3 \ {\rm mJy} \times
                  \left( \frac{T_{\rm ej} }{10^{49}{\rm erg}} \right)
                  \left( \frac{n_0}{\rm cm^{-3}}\right)^{\frac{p+1}{4}} 
                  \left( \frac{\epsilon_B}{\rm 0.1}\right)^{\frac{p+1}{4}} \nonumber \\
                  & \times \left( \frac{\epsilon_e}{\rm 0.1}\right)^{p-1}                  
                  \left( \frac{v_{\rm ej}}{1} \right)^{\frac{5p-7}{2}}
                  \left( \frac{D}{10^{27} {\rm cm}} \right)^{-2}  \left( \frac{\nu_{\rm obs}}{1.4 {\rm GHz}} \right)^{-\frac{p-1}{2}} 
\end{align}   
for an observation frequency $\nu_{\rm obs}$ higher than the 
self-absorption and synchrotron peak frequency at a distance $D$.
The parameters $\epsilon_B$ and $\epsilon_e$, both set to $0.1$,
determine how efficient the energy of the blast wave is transfered to the magnetic field and to
electrons. 
$n_0$ denotes the surrounding particle density 
and is set to 
$0.1 {\rm cm^{-3}}$~\footnote{Notice that the overall uncertainty 
on the density of the surrounding material is rather large. 
To constrain the EOSs or extract the binary parameters 
from radio observations strict bounds on $n_0$ will be needed.}.
Additionally we assume $p=2.3$ and $\nu_{\rm obs}=1.4 {\rm GHz}$, 
as done in~\cite{Nakar:2011cw}.

In figure~\ref{fig:radio} we present for four different EOSs 
the expected peak time $t_{\rm peak}$ (upper panel) and radio fluence 
${F^\nu}^{\rm rad}_{\rm peak}$ (lower panel). 
We find that for an increasing total mass the peak time 
$t_{\rm peak}^{\rm rad}$ decreases while the 
peak fluency ${F^\nu}^{\rm rad}_{\rm peak}$ increases.
For larger mass ratios the peak fluency is largest. 
Considering different EOSs we find significant differences. 
In general the observable peak time in the radio band, 
i.e.~$t_{\rm peak}^{\rm rad}$, happens later for 
softer EOSs, for those setups also the peak fluency is higher.

\section{Conclusion}
\label{sec:conclusion}

\subsection{Summary}
\label{sec:summary}

In this work we have derived fitting functions for the main
ejecta properties from binary neutron star mergers, 
namely the mass, kinetic energy, and velocity of the unbound material. 
Our work is (as a first step) restricted to dynamical 
ejecta for which a large number of numerical simulation data are available.  
In total we use a sample of 172 numerical simulations 
of binary neutron star mergers to derive our fits. 
The high number of data points 
allows to cover a large region of the possible 
binary neutron star parameter 
space including 23 different EOSs, total masses between $2.4 M_\odot$ and $4 M_\odot$, 
and mass ratios between $q=1.0$ and $q\approx2.1$.
The residual errors of the fitting functions are of the order 
of the uncertainty of the numerical relativity results. 

Additionally, we presented estimates for the geometry of the ejected material
and compared those with numerical relativity simulations. 
We found that the high density region of the ejected material can be 
approximated by a three dimensional annular sector, i.e.~a crescent-like structure. 

Using the results of~\cite{Sekiguchi:2016bjd} we also discussed the influence 
of the EOS and mass ratio on the electron fraction inside the ejected material, 
where in general softer and higher mass ratio configurations 
are characterized by lower electron fractions. 
Following~\cite{Dietrich:2016lyp} we presented 
how the intrinsic rotation on the individual neutron 
stars affects the ejecta mass, where we found in particular that for high mass ratios
the aligned spin of the lower star increases the amount of the ejected material.

Based on estimated ejecta properties we studied possible electromagnetic observables 
for binary neutron star mergers. In particular, we have focused on the possibility 
of the formation of kilonovae and radio flares. 
Considering kilonovae, analytical models have been employed to determine the 
time when the kilonovae is brightest as well as the corresponding luminosity and 
temperature. While these estimates just represent the properties of the EM counterpart 
at a fixed time, we also used the model proposed in~\cite{Kawaguchi:2016ana}
to derive the time evolution of the luminosity and light curve. 
We checked the model against radiative transfer simulations of~\cite{Tanaka_web}
and found good agreement.

Finally, we estimated the peak time and peak fluency of the radio flares produced after 
the binary neutron star merger.  
Those flares will be observable month up to years after the merger. 

\subsection{Consequences for future observations}
\label{sec:consequences}

The first two GW detections GW150914 and GW151226 have proven 
that pipelines for EM follow studies are in place and work reliably. 
Detailed informations can be found in~\cite{Abbott:2016gcq} and references therein. 
However, in case of an upcoming GW detection of a BNS system an estimate
about corresponding kilonovae and radio flares may support follow up studies. 

Once a GW is detected the first parameter estimates for the binary properties are produced
within the first minutes after the detection. 
This time is small enough to allow observations in the 
visible, near-infrared, and radio band. 

On a practical term it is important to point out that the time between
the GW detection and the kilonovae observation is too short to perform full NR simulations, 
which typically have run times of the order of weeks to months. 
Thus, once the first knowledge about the properties of the binary is available
phenomenological formulas, as presented here, are needed to obtain estimates for possible EM counterparts. 
After the kilonovae observation NR simulations with microphysical descriptions as 
neutrinos transport, tabulated EOS, and magnetic fields can be performed 
to obtain more reliable results. At this stage, 
our estimates help to reduce the region in the parameter 
space which have to be covered by NR simulations. 

Notice that the situation is different for radio flares, which are detectable 
on the order of years after the merger. Full-NR simulations 
for a variety of parameters can be performed between the detection of the GWs and 
the observation of the radio signal. \\

Overall, our work represents a first step towards a systematic combination between binary parameters 
accessible from gravitational wave observations and electromagnetic counterparts for a large 
range of the binary neutron star parameter space. In the future even more setups have to be included 
testing extreme corners of the parameter space. Furthermore, 
a detailed microphysical description in numerical simulations will help
to account for other effects as e.g.,~magnetic fields and the ejecta 
produced by the disk wind after the formation of the merger remnant.  

\ack
  We thank Sebastiano Bernuzzi, Brett Deaton, Francois Foucart, Kyohei Kawaguchi, 
  Nathan~K.~Johnson-McDaniel, David Radice, Masaomi Tanaka 
  for comments and fruitful discussions. 
  
  It is a pleasure to also thank Matthias Hempel who kindly gave us the 
  EOS tables for cold neutron stars in beta-equilibrium.
  We are grateful to Masaomi Tanaka for making his 
  Monte Carlo simulation data public available and to 
  Kyohei Kawaguchi for making his code to compute lightcurves for 
  BNS systems available. 
  
  Parts of the presented results relied on simulations 
  performed on SuperMUC at the LRZ (Munich) under 
  the project number pr48pu, Jureca (J\"ulich) 
  under the project number HPO21, Stampede 
  (Texas, XSEDE allocation - TG-PHY140019).



\section*{References} 
\bibliographystyle{iopart-num}
\bibliography{bns_ejfits}

\providecommand{\newblock}{}
\begin{thebibliography}{10}
\expandafter\ifx\csname url\endcsname\relax
  \def\url#1{{\tt #1}}\fi
\expandafter\ifx\csname urlprefix\endcsname\relax\def\urlprefix{URL }\fi
\providecommand{\eprint}[2][]{\url{#2}}

\bibitem{Abbott:2016blz}
Abbott B~P {\em et~al.\/} (Virgo, LIGO Scientific) 2016 {\em Phys. Rev.
  Lett.\/} {\bf 116} 061102 (\textit{Preprint} \eprint{1602.03837})

\bibitem{Abbott:2016nmj}
Abbott B {\em et~al.\/} (Virgo, LIGO Scientific) 2016 {\em Phys. Rev. Lett.\/}
  {\bf 116} 241103 (\textit{Preprint} \eprint{1606.04855})

\bibitem{Aasi:2013wya}
Aasi J {\em et~al.\/} (LIGO Scientific Collaboration, Virgo Collaboration) 2016
  {\em Living Rev. Relativity\/} {\bf 19} 1 (\textit{Preprint}
  \eprint{1304.0670})

\bibitem{Abbott:2016ymx}
Abbott B~P {\em et~al.\/} (Virgo, LIGO Scientific) 2016  (\textit{Preprint}
  \eprint{1607.07456})

\bibitem{Paczynski:1986px}
Paczynski B 1986 {\em Astrophys. J.\/} {\bf 308} L43--L46

\bibitem{Eichler:1989ve}
Eichler D, Livio M, Piran T and Schramm D~N 1989 {\em Nature\/} {\bf 340}
  126--128

\bibitem{Soderberg:2006bn}
Soderberg A~M {\em et~al.\/} 2006 {\em Astrophys. J.\/} {\bf 650} 261--271
  (\textit{Preprint} \eprint{astro-ph/0601455})

\bibitem{Tanvir:2013pia}
Tanvir N, Levan A, Fruchter A, Hjorth J, Wiersema K {\em et~al.\/} 2013 {\em
  Nature\/} {\bf 500} 547 (\textit{Preprint} \eprint{1306.4971})

\bibitem{Yang:2015pha}
Yang B, Jin Z~P, Li X, Covino S, Zheng X~Z, Hotokezaka K, Fan Y~Z, Piran T and
  Wei D~M 2015 {\em Nature Commun.\/} {\bf 6} 7323 (\textit{Preprint}
  \eprint{1503.07761})

\bibitem{Jin:2016pnm}
Jin Z~P, Hotokezaka K, Li X, Tanaka M, D'Avanzo P, Fan Y~Z, Covino S, Wei D~M
  and Piran T 2016  (\textit{Preprint} \eprint{1603.07869})

\bibitem{Nakar:2011cw}
Nakar E and Piran T 2011 {\em Nature\/} {\bf 478} 82--84 (\textit{Preprint}
  \eprint{1102.1020})

\bibitem{Goriely:2011vg}
Goriely S, Bauswein A and Janka H~T 2011 {\em Astrophys.J.\/} {\bf 738} L32
  (\textit{Preprint} \eprint{1107.0899})

\bibitem{Rosswog:2013kqa}
Rosswog S, Korobkin O, Arcones A, Thielemann F~K and Piran T 2014 {\em Mon.
  Not. Roy. Astron. Soc.\/} {\bf 439} 744--756 (\textit{Preprint}
  \eprint{1307.2939})

\bibitem{Grossman:2013lqa}
Grossman D, Korobkin O, Rosswog S and Piran T 2014 {\em Mon. Not. Roy. Astron.
  Soc.\/} {\bf 439} 757--770 (\textit{Preprint} \eprint{1307.2943})

\bibitem{Tanaka:2013ana}
Tanaka M and Hotokezaka K 2013 {\em Astrophys.J.\/} {\bf 775} 113
  (\textit{Preprint} \eprint{1306.3742})

\bibitem{Hotokezaka:2013kza}
Hotokezaka K, Kyutoku K, Tanaka M, Kiuchi K, Sekiguchi Y, Shibata M and Wanajo
  S 2013 {\em Astrophys. J.\/} {\bf 778} L16 (\textit{Preprint}
  \eprint{1310.1623})

\bibitem{Faber:2012rw}
Faber J~A and Rasio F~A 2012 {\em Living Rev.Rel.\/} {\bf 15} 8
  (\textit{Preprint} \eprint{1204.3858})

\bibitem{Baiotti:2016qnr}
Baiotti L and Rezzolla L 2016  (\textit{Preprint} \eprint{1607.03540})

\bibitem{Hotokezaka:2012ze}
Hotokezaka K, Kiuchi K, Kyutoku K, Okawa H, Sekiguchi Y~i {\em et~al.\/} 2013
  {\em Phys.Rev.\/} {\bf D87} 024001 (\textit{Preprint} \eprint{1212.0905})

\bibitem{Bauswein:2013yna}
Bauswein A, Goriely S and Janka H~T 2013 {\em Astrophys.J.\/} {\bf 773} 78
  (\textit{Preprint} \eprint{1302.6530})

\bibitem{Dietrich:2015iva}
Dietrich T, Bernuzzi S, Ujevic M and Br{\"u}gmann B 2015 {\em Phys. Rev.\/}
  {\bf D91} 124041 (\textit{Preprint} \eprint{1504.01266})

\bibitem{Lehner:2016lxy}
Lehner L, Liebling S~L, Palenzuela C, Caballero O~L, O'Connor E, Anderson M and
  Neilsen D 2016  (\textit{Preprint} \eprint{1603.00501})

\bibitem{Sekiguchi:2016bjd}
Sekiguchi Y, Kiuchi K, Kyutoku K, Shibata M and Taniguchi K 2016 {\em Phys.
  Rev.\/} {\bf D93} 124046 (\textit{Preprint} \eprint{1603.01918})

\bibitem{Dietrich:2016hky}
Dietrich T, Ujevic M, Tichy W, Bernuzzi S and Bruegmann B 2016
  (\textit{Preprint} \eprint{1607.06636})

\bibitem{Metzger:2016pju}
Metzger B~D 2016  (\textit{Preprint} \eprint{1610.09381})

\bibitem{Read:2009yp}
Read J~S, Markakis C, Shibata M, Uryu K, Creighton J~D {\em et~al.\/} 2009 {\em
  Phys.Rev.\/} {\bf D79} 124033 (\textit{Preprint} \eprint{0901.3258})

\bibitem{Foucart:2012nc}
Foucart F 2012 {\em Phys. Rev.\/} {\bf D86} 124007 (\textit{Preprint}
  \eprint{1207.6304})

\bibitem{Kawaguchi:2016ana}
Kawaguchi K, Kyutoku K, Shibata M and Tanaka M 2016 {\em Astrophys. J.\/} {\bf
  825} 52 (\textit{Preprint} \eprint{1601.07711})

\bibitem{Yagi:2016bkt}
Yagi K and Yunes N 2016  (\textit{Preprint} \eprint{1608.02582})

\bibitem{Rosswog:2015nja}
Rosswog S 2015 {\em Int.J.Mod.Phys.\/} {\bf D24} 1530012 (\textit{Preprint}
  \eprint{1501.02081})

\bibitem{Radice:2016dwd}
Radice D, Galeazzi F, Lippuner J, Roberts L~F, Ott C~D and Rezzolla L 2016 {\em
  Mon. Not. Roy. Astron. Soc.\/} {\bf 460} 3255--3271 (\textit{Preprint}
  \eprint{1601.02426})

\bibitem{Palenzuela:2015dqa}
Palenzuela C, Liebling S~L, Neilsen D, Lehner L, Caballero O~L, O'Connor E and
  Anderson M 2015 {\em Phys. Rev. D\/} {\bf 92} 044045 (\textit{Preprint}
  \eprint{1505.01607})

\bibitem{Dietrich:2016lyp}
Dietrich T, Bernuzzi S, Ujevic M and Tichy W 2016  (\textit{Preprint}
  \eprint{1611.07367})

\bibitem{Li:1998bw}
Li L~X and Paczynski B 1998 {\em Astrophys.J.\/} {\bf 507} L59
  (\textit{Preprint} \eprint{astro-ph/9807272})

\bibitem{Metzger:2010sy}
Metzger B, Martinez-Pinedo G, Darbha S, Quataert E, Arcones A {\em et~al.\/}
  2010 {\em Mon.Not.Roy.Astron.Soc.\/} {\bf 406} 2650 (\textit{Preprint}
  \eprint{1001.5029})

\bibitem{Roberts:2011xz}
Roberts L~F, Kasen D, Lee W~H and Ramirez-Ruiz E 2011 {\em Astrophys.J.\/} {\bf
  736} L21 (\textit{Preprint} \eprint{1104.5504})

\bibitem{Korobkin:2012uy}
Korobkin O, Rosswog S, Arcones A and Winteler C 2012  (\textit{Preprint}
  \eprint{1206.2379})

\bibitem{Tanaka:2016sbx}
Tanaka M 2016 {\em Adv. Astron.\/} {\bf 2016} 6341974 (\textit{Preprint}
  \eprint{1605.07235})

\bibitem{Rosswog:2016dhy}
Rosswog S, Feindt U, Korobkin O, Wu M~R, Sollerman J, Goobar A and
  Martinez-Pinedo G 2016  (\textit{Preprint} \eprint{1611.09822})

\bibitem{Fernandez:2015use}
Fernández R and Metzger B~D 2016 {\em Ann. Rev. Nucl. Part. Sci.\/} {\bf 66}
  2115 (\textit{Preprint} \eprint{1512.05435})

\bibitem{Kasen:2013xka}
Kasen D, Badnell N~R and Barnes J 2013 {\em Astrophys. J.\/} {\bf 774} 25
  (\textit{Preprint} \eprint{1303.5788})

\bibitem{Wanajo:2014wha}
Wanajo S, Sekiguchi Y, Nishimura N, Kiuchi K, Kyutoku K and Shibata M 2014 {\em
  Astrophys. J.\/} {\bf 789} L39 (\textit{Preprint} \eprint{1402.7317})

\bibitem{Tanaka:2013ixa}
Tanaka M, Hotokezaka K, Kyutoku K, Wanajo S, Kiuchi K, Sekiguchi Y and Shibata
  M 2014 {\em Astrophys. J.\/} {\bf 780} 31 (\textit{Preprint}
  \eprint{1310.2774})

\bibitem{Tanaka_web}
{Webpage Tanaka}
  \urlprefix\url{{http://th.nao.ac.jp/MEMBER/tanaka/nr_merger_lightcurve.html}}

\bibitem{Kawaguchi_web}
{Webpage Karaguchi}
  \urlprefix\url{{http://www2.yukawa.kyoto-u.ac.jp/~kyohei.kawaguchi/kn_calc/main.html}}

\bibitem{Abbott:2016gcq}
Abbott B~P {\em et~al.\/} (InterPlanetary Network, DES, INTEGRAL, La
  Silla-QUEST Survey, MWA, Fermi-LAT, J-GEM, DEC, GRAWITA, Pi of the Sky, Fermi
  GBM, MASTER, Swift, iPTF, VISTA, ASKAP, SkyMapper, PESSTO, TOROS, Pan-STARRS,
  Virgo, Liverpool Telescope, BOOTES, LIGO Scientific, LOFAR, C2PU, MAXI) 2016
  {\em Astrophys. J.\/} {\bf 826} L13 (\textit{Preprint} \eprint{1602.08492})

\end{thebibliography}


\end{document}